\documentclass[aps,floats,prb,showpacs,twocolumn]{revtex4}

\newcommand{\beq}{\begin{eqnarray}} \newcommand{\eeq}{\end{eqnarray}}

\def\eq{&=&} \def\nn{\nonumber} \def\tr{{\,\rm tr\,}} \def\be{{\bf e}}
\def\bl{{\bf l}} \def\bp{{\bf p}} \def\bq{{\bf q}} \def\bx{{\bf x}}
\def\cG{{\cal G}} \def\cN{{\cal N}} \def\cT{{\cal T}}  \def\cZ{{\cal Z}} 

\usepackage{amsfonts,amsmath} \usepackage{bm} \usepackage{dcolumn}
\usepackage{epsfig} \usepackage{latexsym}

\begin{document}

\title{Inelastic electron transport in granular arrays}

\author{Alexander Altland$^1$, Leonid I. Glazman$^{2,3}$, Alex
  Kamenev$^2$, and Julia S. Meyer$^4$}

\affiliation{$^1$Institut f\"ur
  Theoretische Physik, Universit\"at zu K\"oln, 50937 K\"oln, Germany\\
  $^2$Department of Physics, and $^3$W.I. Fine Theoretical
  Physics Institute,\\
  University of Minnesota, Minneapolis, MN55455, USA\\
  $^4$Materials Science Division, Argonne National Laboratory,
  Argonne, IL60439, USA,\\
  and Department of Physics, The Ohio State University, Columbus,
  OH43210, USA}

\date{\today}


\begin{abstract}
  Transport properties of granular systems are governed by Coulomb
  blockade effects caused by the discreteness of the electron charge.
  We show that, in the limit of vanishing mean level spacing on the
  grains, the low--temperature behavior of 1{{d}} and 2{{d}} arrays is
  insulating at any inter--grain coupling (characterized by
  a dimensionless conductance $g$.) In 2{{d}} and $g\gg 1$, there is a
  sharp Berezinskii--Kosterlitz--Thouless crossover to the conducting
  phase at a certain temperature, $T_{\rm BKT}$. These results are
  obtained by applying an instanton analysis to map the conventional  `phase'
  description of  granular arrays onto the dual `charge'
  representation.
\end{abstract}

\maketitle
\section{Introduction}
\label{sec-intro}

It has been long appreciated that the low--temperature physics of
generic disordered metals is characterized by a subtle interplay
of  electron--electron interactions and  coherent disorder
scattering. While both effects are of crucial importance, their
unified treatment still evades a complete theoretical description.
It is useful, thus, to approach them separately. The limiting case
of ``coherence without interactions''   has been studied
intensely. It is well understood that the coherent multiple
scattering off impurities leads to Anderson localization: in one and two
dimensions all states are localized~\cite{anderson}. While in
homogeneously disordered systems this phenomenon has always to be taken into
account, granular systems admit for a parameter regime where the
physics  is entirely controlled by  interaction effects. It is the
purpose of this paper to explore the regime of ``interactions
without coherence'' accessible in metallic granular arrays.

Metallic  granular arrays are also of great interest in their own
right~\cite{Abeles,fs2,Schon90,Gerber97,Jager,Golubev92,tschersich,Beloborodov01,AGK,MKG,vinokur,vinokur2,Tripathi,shklovskii}.
In particular, strongly coupled arrays ($g\gg1$, where $g$  is the
dimensionless inter--grain conductance) have become a subject of
increased theoretical attention in recent
years~\cite{tschersich,Beloborodov01,AGK,MKG,vinokur,vinokur2,Tripathi}.
An isolated grain or quantum dot is characterized by three energy
scales:  the Thouless energy $E_{\rm Th}$, the charging energy
$E_c$, and the mean level spacing $\delta$. The tunneling
coupling between the grains adds another parameter: the
dimensionless conductance $g$ of a contact between two neighboring
grains. Throughout this paper we focus on the regime, where the
Thouless energy $E_{\rm Th}$ is the largest energy scale. This
allows one to treat each grain as a zero--dimensional object and
disregard the intra--grain  in comparison with the inter--grain
resistance. The interaction effects are controlled by the charging
energy $E_c$ (in the simplest model $E_c=e^2/(2C)$, where $C$ is
the self--capacitance of a grain.) In our studies it is the next
largest energy scale in the system. Finally, quantum coherence
effects are governed by the energy scale(s) $\propto \delta$. If
such a scale is much smaller than all relevant temperatures,  one
may treat each grain as having a {\em continuous} spectrum. This
assumption allows one to disregard phase coherence. In essence: an
electron exiting a grain is never the same electron that has
previously entered it.

The parameter regime specified above justifies the ``interactions
without coherence'' program. It is clear though, that such a
simplification cannot work down to the very smallest temperatures. At
low enough temperature, coherent propagation through multiple grains
will become important and our approximation is bound to fail. It is,
thus, important to realize that the subject of our considerations is a
transient (though possibly wide) temperature range.  In this range,
coherence may be disregarded while interactions (and inter--grain
tunneling) are crucially important in determining the electrical
properties of the array.

At small inter--grain conductance $g\ll 1$, an electron is completely
localized within a single grain. Therefore, the problem is reduced to the
description of {\em classical} charges moving on the lattice (which
is a simple limiting case of the considerations given below.) In
the simplest case of on--site interactions only, there is an
energy barrier $E_c$ impeding the transition of electrons between two
neighboring grains. It is thus natural to expect  activation
behavior of the conductivity, with the activation temperature
$T^*=E_c$.

The present paper is devoted to the  more intricate case of
 large inter--grain conductances, $g\gg 1$. In this case the
charge may spread over many grains to decrease its charging
energy.  The interplay of interactions and tunneling dictates
that this spreading involves an (exponentially) large, but {\em
finite} number of grains. As a result,  the lowest energy
excitations of the system are large single--charge {\em solitons}.
The activation energy for creating such an extended charge carrier
is substantially reduced, leading to the low--temperature
conductivity of the form:
\begin{equation}
\label{eq-res} \sigma(T) =g\,\exp\left[-\frac{T^*}T\;\right]\,,
\end{equation}
where $T^*_{(1{{d}})}\sim gE_c\exp[-g/4]$ and $T^*_{(2{{d}})}\sim
g^2E_c\exp[-g]$. The important consequence of Eq.~(\ref{eq-res}) is
that one-- and two--dimensional arrays are insulating at arbitrarily
large inter--grain conductance, $g$. This is a pure interaction
effect; Anderson localization physics is {\em not} included in the
model. Switching off the interactions, one obtains Ohmic metallic
behavior with a temperature--independent conductivity.

The solitons interact with each other up to distances comparable to
their (exponentially large) radius, even if the initial model
possesses on--site interactions only. Once they start to overlap,
Eq.~(\ref{eq-res}) is not valid anymore. In 1{{d}} this happens at $T\sim
T^*_{(1{{d}})}$, where the conductivity smoothly crosses over to its
high--temperature behavior~\cite{tschersich} $\sigma_{(1{{d}})}(T)=g-2\ln
(gE_c/T)$. In two dimensions, the solitons interact logarithmically
over a large range of distances. This leads to a
Berezinskii--Kosterlitz--Thouless (BKT) unbinding of
soliton--anti-soliton pairs~\cite{BKT-b,BKT-kt} at the temperature
\begin{equation}
\label{eq-resBKT} T_{\rm BKT}=T^*_{(2{{d}})}/g \ll T^*_{(2{{d}})}   \, .
\end{equation}
Around this temperature the conductivity undergoes a sharp
crossover 
from the exponentially small value given by $\sigma(T_{\rm BKT})$,
cf.~Eq.~(\ref{eq-res}), up to the high--temperature asymptotics,
$\sigma_{(2{{d}})}(T)=g-\ln (gE_c/T)$. In the model with only mutual
capacitances between neighboring grains the Coulomb interaction is
logarithmic at arbitrarily large distances. This results in a true
BKT phase transition with zero conductivity below the transition
point. The $g\ll 1$ version of the latter model was previously
considered in Refs.~[\onlinecite{fs2,fs}]. The introduction of
on--grain Coulomb interactions transforms the transition into a
crossover. Interestingly, for $g\gg 1$ the BKT remains sharp even for
the pure on--grain (self--capacitance only) Coulomb interactions.

Technically we approach the problem from two complimentary
perspectives: the {\em phase} and the {\em charge}
representations. The former is straightforwardly derived from the
microscopic fermionic model~\cite{AES}. It is commonly employed in the
study of both homogenous and granular interacting systems. While being
effective in the high--temperature regime, it becomes increasingly
difficult to handle at lower temperatures. To treat this latter
regime, we employ the charge model, introduced previously within the
context of quantum dot physics~\cite{flensberg,matveev}. Our main
technical achievement is the proof of equivalence of these two
approaches over a parametrically wide range of temperatures. For these
temperatures, both models may be handled in a {\em controlled} way. We
thus conclude that the charge model, although not directly deduced
from the microscopic Hamiltonian, is indeed the proper description of
the low--temperature phase of the system. The results mentioned above
(as well as others discussed below) then follow in an almost
straightforward manner from the charge description.

The equivalence of the two approaches is based on a very important
observation. The charge discreteness (crucial in the
low--temperature insulating phase) manifests itself in the phase
model through the $2\pi$--periodicity of the phase field (the
internal space of the field is the circle $S^1$.) The latter
results in the existence of topologically distinct
stationary--point field configurations,  classified by the integer
winding numbers $W_{\bf l}$ (where the vector index ${\bf l}$
numerates the grains on the lattice.) In  strongly connected
arrays, $g\gg 1$, the action cost for  configurations with
non--zero winding numbers (so--called Korshunov
instantons~\cite{Korshunov87}) is exponentially large. However,
one has to take into account Gaussian fluctuations around the
topologically non--trivial stationary points, which yield a factor
$(gE_c/T)^{1/d}$ for each winding number mismatch between
neighboring grains. This factor suggests that the instanton
configurations  are increasingly important at low
temperatures~\cite{foot1}.  Summation of the  instanton ``gas''
along with the corresponding Gaussian fluctuations and the
phase--volume factors results exactly in the classical
(low--frequency)
 limit of the $d$--dimensional charge model. Specifically
(see below),  the instanton expansion of the phase model coincides
term by term with the perturbative expansion in  back--scattering
amplitudes of the charge model. Therefore, we are convinced  that
the explicit account for  instantons in the phase--like models
is  imperative to restore the charge discreteness and, thus, to
describe the insulating phase.

One may justifiably worry about the role of non--Gaussian
fluctuations. The latter are known to become large at a low enough
temperature $T_0\sim E_c e^{-dg/2}$, violating the validity of the
instanton gas picture.
Crucially, however, (in 1{{d}} and 2{{d}}) the
corresponding charge model predicts an activation gap which is {\em
  parametrically larger} (exponential (in 
$g$) in 1{{d}} and algebraic in 2{{d}}) than
$T_0$.
As a result, there is a wide range of temperatures, where the
fluctuations are well under control, while the physics is completely
dominated by the proliferation of instantons.
The latter results in the appearance of the unit--charge extended
solitons as low--energy charged excitations and, thus, in
activation insulating behavior, Eq.~(\ref{eq-res}).  In 3{{d}},
proliferation of instantons and the onset of strong non--Gaussian
fluctuations, resp., take place at comparable temperatures. As a
result the phase--charge equivalence cannot be reliably
established. It seems plausible, however, that the instanton gas
--- and thus the corresponding charge representation --- provide a
qualitatively correct description of the 3{{d}} insulator as well.

This paper is an extension of two previous shorter
publications~[\onlinecite{AGK,MKG}].  Its intent is
two--fold. Firstly, we present some new results. In particular, we
extend calculations beyond the tunneling limit, accounting for
arbitrary transmission amplitudes between neighboring
grains. Furthermore, in addition to an evaluation of the transport
properties, we discuss the behavior of the single--particle density of
states (DoS). Secondly, we bring out the philosophy of our approach and
expose extensive technical details of the calculations. Our main
message is that charge quantization is crucial in describing the
low--temperature physics of the array --- and, therefore, a
description in terms of charge degrees of freedom is appropriate. As
mentioned above, this description is obtained by accounting for
topologically non--trivial field configurations in the phase
picture. This goes beyond the commonly used perturbative treatment of
the phase model. In 1{{d}} and 2{{d}} arrays, the latter completely misses the
appearance of a new temperature scale $T^*$ marking the crossover to
insulating behavior.

The paper is organized as follows: in Sec.~\ref{sec-models}, we
introduce the phase and charge models. Before coming to the main part,
namely quantum dot arrays, in Sec.~\ref{sec-1dot}, we discuss the
physics of a single dot connected to two
leads. Sec.~\ref{sec-array-1d} discusses one--dimensional arrays
whereas Sec.~\ref{sec-array-2d} contains the two--dimensional
arrays. The conclusion and open questions are discussed in
Sec.~\ref{sec-conclude}.


\section{Phase and Charge Representations}
\label{sec-models}

In this section, we introduce two effective models used to describe
$d$--dimensional quantum dot arrays. As mentioned in the introduction,
the two descriptions are optimally adjusted to the nominally metallic
and the nearly insulating regime, respectively. The application of the
two models to the computation of observables, and the mapping of one
onto the other will be discussed in later sections.

Widely used in the literature is the so--called
Ambegoakar--Eckern--Sch{\"o}n (AES) model~\cite{AES,Schon90} --- a
description of arrays in terms of phase fields. In the limit of
vanishing level spacing, this model may be derived starting from a
microscopic description in terms of electronic degrees of freedom. The
model is presented in Sec.~\ref{subsec-phase}, and its derivation is
reviewed in App.~\ref{app-AES}.

While the AES approach provides an efficient description of the
high--temperature regime, it is untractable in the low--temperature
regime where interaction effects become significant. Rather, at low
temperatures, an alternative description in terms of charge degrees of
freedom is more appropriate. This latter formulation may be derived
from a phenomenological model introduced by Flensberg~\cite{flensberg}
and Matveev~\cite{matveev}. We review the derivation in
Sec.~\ref{subsec-charge}.

The equivalence of the two models --- established by a mapping between
them --- will be discussed at later stages.

\subsection{Phase model}
\label{subsec-phase}

In the regime, where the level spacing of the dot is negligible, the
dot can be described by a single degree of freedom. Starting from a
description in terms of electrons, a phase field $\phi$ is introduced
to decouple the interaction on the dot. Subsequently the electronic
degrees of freedom can be integrated out, yielding an effective theory
in terms of $\phi$. The time--derivative of $\phi$ corresponds to the
voltage $V$ on the dot: $V(\tau)=\dot\phi(\tau)$, where $\tau$ is
imaginary time.

Since the AES model is largely standard by now, we here restrict
ourselves to a brief discussion of its main elements. (For an outline
of its derivation, see App.~\ref{app-AES}.) The phase action $S$
consists of two terms, $S=S_c+S_t$, describing the charging
interaction on the grains and the tunneling between neighboring
grains, respectively.  For the $d$--dimensional array geometry, the
charging term reads
\begin{eqnarray}
\label{eq-S_c}
S_c[\phi]=\sum_\bl\int d\tau
\,\left(\frac{\dot\phi_\bl^2}{4E_c}-iq\dot\phi_\bl\right),
\end{eqnarray}
where $\bl$ is a $d$--dimensional index, denoting the position of the
grain.  Here, $E_c=e^2/(2C)$ is the charging energy, where $e$ is the
electronic charge and $C$ the self--capacitance of the dot. The
dimensionless quantity $q=V_{\rm g}C/e$ is the background charge on
the dot as determined by an external gate voltage $V_{\rm g}$. The
phase fields $\phi_\bl$ obey the boundary condition
$\phi_\bl(\beta)-\phi_\bl(0)=2\pi W$ (where $W\in\Bbb{Z}$.)

The tunneling term is given by:
\begin{equation}
\label{eq-AES}
S_t[\phi]=
-\frac1{16}\sum_{\langle\bl,\bl'\rangle} \sum_k\kappa_k
\tr\left[\left(\Lambda
    e^{i\phi_{\bl\bl'}}\Lambda e^{-i\phi_{\bl\bl'}}\right)^k\right],
\end{equation}
where $\phi_{\bl\bl'}=\phi_\bl-\phi_{\bl'}$. The matrix $\Lambda$
takes the form $\Lambda_{nm}=\delta_{nm}\,{\rm
sign}\,(\epsilon_n)$ in Matsubara basis
($\epsilon_n=2\pi(n\!+\!1/2)T$). Furthermore, the coefficients
$\kappa_k$ are related to the tunneling matrix elements $T_\alpha$
and the density of states $\nu$ as
$\kappa_k=-4\frac{(-1)^k}k\sum_\alpha|\pi\nu T_\alpha|^{2k}$.
(Note that we do not need to require the transmission in every
channel, $\alpha$, to be small --- tunneling can be taken into
account to arbitrary order~\cite{tunneling}.)

Having presented an effective action for the phase field $\phi$, we
proceed by discussing its properties. A large inter--grain conductance
$g$ suppresses dynamical phase fluctuations: in a conductor, voltage
fluctuations are small.  As a first step, one may, thus, expand the
action up to second order in $\phi$. The quadratic tunneling action
reads
\begin{equation}
\label{eq-ohm}
S_t^{(2)}[\phi]=\frac {gT}{4\pi}
\sum_{\langle\bl,\bl'\rangle,m}|\omega_m|\phi_{\bl\bl',m}^{\enspace
  2}.
\end{equation}
Here, the dimensionless conductance of the contacts, $g$, is given
by~\cite{tunneling}
\begin{eqnarray}
g \eq \sum_k k^2\kappa_k =\sum_\alpha\cT_\alpha,
\end{eqnarray}
where $\cT_\alpha=4\pi^2\nu^2
|T_\alpha|^2/(1+\pi^2\nu^2|T_\alpha|^2)^2$ is the transmission
probability in the channel $\alpha$.  The action (\ref{eq-ohm})
describes Ohmic dissipation. Evaluating transport properties of the
array within this approximation, one obtains the classical Kirchhoff
laws.

Going beyond the quadratic approximation~\cite{tschersich}, however,
one finds that the conductance is renormalized to smaller values upon
lowering the temperature. Taking into account the terms quartic in $\phi$, one
obtains the renormalized inter--grain conductance,~\cite{beta}
\begin{equation}
\label{eq-efetov}
g\to g-\frac2{dg}\sum_\alpha \cT_\alpha(1-\cT_\alpha)\ln\frac{gE_c}T.
\end{equation}
In the tunneling limit ($\cT_\alpha\ll1$), Eq.~(\ref{eq-efetov})
reduces to~\cite{tschersich} $g(T)=g-(2/d)\ln(gE_c/T)$.

Eq.~(\ref{eq-efetov}) states that in all dimensions interactions
generate logarithmic corrections to the inter--grain conductance
$g$. This result holds as long as the corrections are small.
Perturbation theory breaks down at the temperature where the
corrections become of order of the bare conductance or, in other
words, the renormalized conductance $g(T)$ reaches values of order
1. This defines a temperature scale $T_0\sim E_ce^{-dg/2}$. The
temperature range below this scale is beyond the applicability of
the perturbative treatment.

However, there is more to be extracted from the AES model even at
$T\gg T_0$. Let us return to the full action Eqs.~(\ref{eq-S_c}) and
(\ref{eq-AES}). The phase field $\phi_\bl$ is a periodic variable
$\phi_\bl(\beta)-\phi_\bl(0)=2\pi W$.  Consequently, the conjugate
variable --- which is charge --- is quantized. By using the
perturbative expansion in $\phi_\bl$ around the minimum $\phi_\bl=0$,
all information about periodicity and, therefore, about charge
quantization is lost. Although for $g\gg 1$ phase fluctuations are
heavily suppressed, there are $\phi_\bl$--configurations that explore
this periodicity and, thus, incorporate the manifestations of charge
quantization. As we show in this paper, these phase {\it
  instantons}~\cite{Korshunov87} provide the key to access the
insulating phase of arrays.

Before discussing the instanton physics in single quantum dots as well
as one-- and two--dimensional arrays, we introduce the aforementioned
alternative model to describe the system.

\subsection{Charge model}
\label{subsec-charge}

In this section, we introduce a phenomenological
model describing  the system in terms of charge degrees of freedom,
i.e., those degrees of freedom that become approximately conserved in
near insulating regimes and, therefore, are optimally suited to
describe the low--temperature physics of the system.

\begin{figure}[h]
  \centerline{\epsfxsize=2in\epsfbox{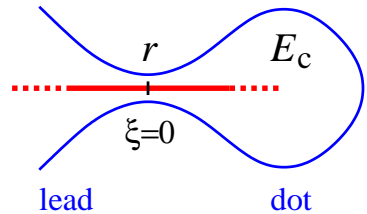}}
\caption{(Color online) Schematic representation of the  charge model. The line
in the middle indicates a clean (infinite) 1{{d}} channel with a
single impurity, having backscattering amplitude $r$.}
\label{fig1}
\end{figure}

Let us start by considering a point contact between a quantum dot and
a metallic reservoir. Due to size quantization effects, no more than a
few transverse modes $\alpha=1,\dots, N$ are permitted to transport
charge across the contact. Each of these modes may be thought of as a
one--dimensional electron liquid. For simplicity, we focus on the case
of just one propagating mode $N=1$ throughout. The generalization to
multi--mode contacts --- essential in order to describe the case of
large dimensionless inter--grain conductance $g\gg1$ --- is discussed
in appendix~\ref{app-gamma}.

Bosonizing the one--dimensional electron liquid in the conventional
way~\cite{flensberg,matveev}, the system is described in terms of a
bosonic field $\theta(\tau,\xi)$. The gradient of this field,
$\partial_\xi \theta(\tau,\xi)$, defines the local electron density,
i.e., the electron number on the dot may be written as
$\cN=\int_0^\infty\! d\xi\, \partial_\xi \theta (\tau,\xi) =-
\theta(\tau,0)$.  This implies that the Coulomb energy takes the
simple form $(e\cN)^2/(2C) = E_c\theta^2 (\tau,0)$. Finally,
accounting for backscattering by introducing a point scatterer of
reflection amplitude $r$ at coordinate $\xi=0$ (cf.~Fig.~\ref{fig1}),
the imaginary--time action of the bosonic field reads
\begin{eqnarray}
                                                    \label{eq-dot}
S[\theta(\tau,z)] &\!=\!&\int\limits_0^\beta d\tau\;
  \Big\{ \int\limits_{-\infty}^{\infty}d\xi
  \left[(\partial_\tau\theta)^2+(\partial_\xi\theta)^2\right]+
   \nonumber \\
  && +  E_c\theta^2(\tau,0) -
\frac{Dr}\pi\cos[2\pi\theta(\tau,0)]\Big\},
\end{eqnarray}
where $D$ is the electronic bandwidth. Integrating over the fields
$\theta(\xi\neq0)$ --- which is possible because their action is
quadratic --- we obtain
\begin{equation}
                                               \label{eq-dot1}
S[\theta] = \frac1T\sum_m \left( \pi|\omega_m|+
E_{\rm c}\right)\theta_{m}^2 - \frac{Dr}\pi\!\int\limits_0^\beta
\!d\tau\,\cos(2\pi\theta(\tau))
\end{equation}
as the effective action of a single remaining degree of freedom
$\theta(\tau)\equiv \theta(\tau,0)$.  Here we have introduced the
Matsubara representation $\theta_m=\int_0^\beta d\tau\,
\theta(\tau)e^{-i\omega_m\tau}$, where $\omega_m=2\pi Tm$. The
dissipative term, $\pi|\omega_m|\theta_{m}^{2}$, appears as a
consequence of the assumption that the mean level spacing is the
smallest energy scale in the model, $\delta\to 0$. It is generated by
integrating out the continuum spectrum of the degrees of freedom on
the dot.

The above expression Eq.~(\ref{eq-dot1}) can be easily generalized
to the array  geometry. A field $\theta_{i,{\bf l}}$ is assigned
to each contact, where the $d$--component index $\bl$ denotes its
position within the array and $i=1\dots d$ labels its direction.
In this notation, the instantaneous electron density on the grain
${\bf l}$ is given by the lattice divergence, $\cN_{\bf
l}=\sum_i(\theta_{i,\bl+\be_i}-\theta_{i,\bl})
=\nabla\cdot\vec\theta_{{\bf l}}$ (where $\be_i$ is a unit vector
in $i$--direction and the vector notation $\vec\theta_{\bf l}$ is
introduced.) The generalization of Eq.~(\ref{eq-dot1}) reads:
\begin{widetext}
\begin{equation}
\label{eq-matveev} S\!\left[\,\vec\theta_{\bf l}\,\right] =
\sum_{\bf l} \left\{\frac1T\sum_m \left(
\pi|\omega_m|\vec\theta_{{\bf l},m}^{\;2} + E_{\rm
c}(\nabla\cdot\vec\theta_{{\bf l},m}-q\,\delta_{m,0})^2 \right) -
\frac{Dr}\pi\sum_{i}\int\limits_0^\beta d\tau\;
\cos(2\pi\theta_{i,{\bf l}}(\tau))\right\},
\end{equation}
\end{widetext}
where $D$ is again the bandwidth. As in the single dot expression
Eq.~(\ref{eq-dot1}), the first term in the action~(\ref{eq-matveev})
describes the dissipative dynamics originating from integrating out
degrees of freedom within the grains, the second term is responsible
for the interaction effects, i.e., charging, and the third one
describes backscattering in the contacts.  Furthermore, we introduced
the external gate voltage, $q$, as an additional control parameter.

As shown in appendix~\ref{app-gamma}, generalization to the
$N$--channel case amounts to replacing the single reflection
coefficient $r$ by the product $\prod_{\alpha=1}^N r_\alpha$,
where $r_\alpha$ is the reflection coefficient in channel
$\alpha$.  We define the dimensionless parameter $\cG_0$ as
\begin{equation}
\label{eq-cg0}
\cG_0=-\sum_\alpha \ln|r_\alpha|^2\enspace\Leftrightarrow\enspace
\prod_\alpha r_\alpha=e^{-\cG_0/2}. 
\end{equation}
Although the single channel expression Eq.~(\ref{eq-matveev}) was
obtained for a small reflection coefficient $r\ll1$, its
multi--channel generalization  remains valid as long as
$\cG_0\gg1$ (i.e., individual reflection coefficients $r_\alpha$
may be arbitrary, $0\!<\!r_\alpha\!<\!1$.)  For a tunneling
contact $\cT_\alpha\ll 1$, where $\cT_\alpha=1-|r_\alpha|^2$ is
the transmission coefficient in channel $\alpha$, this product can
be expressed through the dimensionless conductance
$g=\sum_\alpha\cT_\alpha$ within exponential accuracy as
\begin{equation}
\prod_\alpha
r_\alpha\!=\exp\!\left[\sum_\alpha\ln\!\sqrt{1\!-\!\cT_\alpha}\right]
\!\simeq \exp\!\left[-\frac12\sum_\alpha\cT_\alpha\right]\!=e^{-\frac
  g2}.\nn
\end{equation}
In this regime, the dimensionless  parameter, $\cG_0\simeq g$.

The action (\ref{eq-matveev}) will be our starting point for exploring
interaction effects in single dots as well as in arrays. We will use it
as an alternative to Eqs.~(\ref{eq-S_c}) and (\ref{eq-AES}) and
discuss the connections between the two descriptions in the following sections.


\section{Single quantum dot}
\label{sec-1dot}

The simplest setup on which the impact of interactions on transport
through an almost open system can be studied is a single quantum dot
coupled to two leads~\cite{averin-nazarov,matveev}.  Interesting in
its own right, the discussion of the quantum dot will facilitate the
development of the formalism required to describe arrays. We follow a
three--step program: in Sec.~\ref{subsec-0d_phase} the AES phase model
is investigated, in Sec.~\ref{subsec-0d_charge} the alternative charge
description is used, and in Sec.~\ref{subsec-compare} the two
procedures are compared.

\subsection{Phase model}
\label{subsec-0d_phase}

Consider a single quantum dot coupled to a left ($L$) and right ($R$)
lead. In the limit of a vanishing level spacing $\delta\to0$, the
system may be described by the phase action Eqs.~(\ref{eq-S_c})
and (\ref{eq-AES}),
\begin{eqnarray}
\label{eq-S_0}
\!S= \!\int \!\! d\tau\!
\left(\frac{\dot\phi^2}{4E_c}\!-\!iq\dot\phi\right)
\!+\!\frac1{4}\sum_k\kappa_k
\tr\!\left[\left(\Lambda
    e^{i\phi}\Lambda e^{-i\phi}\right)^k\right]\!.
\end{eqnarray}
The perturbative results one may derive form this action have been
discussed in the previous section. Going beyond this level, we here
include topologically non--trivial excitations and discuss the resulting
charge quantization effects.

In addition to the constant solution $\phi=0$, the tunneling part
of the action is stationary on the so--called Korshunov instanton
configurations~\cite{Korshunov87}.  These additional saddle point
solutions are characterized by their winding number
$W=(\phi(\beta)-\phi(0))/(2\pi)$ and can be represented
as~\cite{Korshunov87,Zaikin91,Grabert96,Nazarov99}
\begin{equation}
\label{Korshunov_inst}
e^{i\phi_W(\{z\},\tau)}=\prod_{a=1}^{|W|}\left[\frac{e^{2\pi i\tau
    T}-z_a}{1-\bar z_ae^{2\pi i\tau T}}\right]^{{\rm sign}\,W}.
\end{equation}
Here the $|W|$ complex parameters $z=(z_1\dots z_{|W|})$ are subject
to the condition $|z_a|<1$.  The temporal variation of $\phi$
corresponds to a voltage pulse $V=i\dot\phi$ on the dot: the
parameters $1-|z|$ determine the duration of the voltage pulse and
arg$\,z$ its instance ($z=0$ corresponds to a linear phase profile or
a constant voltage.)

In the limit $T\ll E_c$, the action is dominated by the tunneling term,
implying that the Korshunov instantons are approximate saddle point
configurations of the total action.  Substituting
Eq.~(\ref{Korshunov_inst}) into the action (\ref{eq-S_0}), one finds
that~\cite{Nazarov99}
\begin{eqnarray}
S[\phi_W(\{z\})]\approx\cG_0|W|-2\pi i Wq,
\end{eqnarray}
where the dimensionless conductance $\cG_0$ is given by
(cf.~Eq.~(\ref{eq-cg0})) $\cG_0=\sum_k\kappa_{2k-1}$. Apart from a
small charging contribution, $S_c[\phi_W(\{z\})] = \pi^2(T/E_c)
\sum_{a,a'} (1-|z_a|^2|z_{a'}|^2) / ((1-z_az_{a'}^*)(1-z_a^*z_{a'}))$,
the action is independent of $z_a$, i.e., the variables $z_a$ are
instanton zero modes.

Due to the largeness of the parameter $\cG_0\gg1$, the
contribution of a single instanton is exponentially small.
However, as will be shown in the following, fluctuations around
the instanton trajectory increase with decreasing
temperature~\cite{Grabert96}, i.e., a temperature scale exists
below which the instanton contributions become important.

The partition function $Z$ can be represented as a sum over different
winding number sectors:
\begin{eqnarray}
\label{eq-sum_W}
Z = Z_0\sum\limits_W \frac{Z_W}{Z_0} e^{2\pi i Wq},
\end{eqnarray}
where $Z_W$ is the contribution from configurations with winding
number $W$. Note that a given total winding number $W$ can be obtained
by superposition of a sequence of $s+W$ instantons and $s$
anti-instantons. Although these configurations are not true saddle
point solutions, it can be shown that the interaction between
instantons is weak~\cite{fkls} and the ideal (instanton) gas
approximation may be used.  Referring for a detailed account of the
computation of the corresponding fluctuation
determinant~\cite{Grabert96} to App.~\ref{app-fluct_det}, we here
sketch the main steps.

Starting from the action (\ref{eq-S_0}), one expands in small
fluctuations $\delta\phi$ around the instanton configuration
$\phi_W(\{z\})$. We denote the Gaussian fluctuation contribution to
the action by
\begin{eqnarray}
\delta S_{\rm inst}=g\langle\delta\phi|\hat
  F_W|\delta\phi\rangle,\nn
\end{eqnarray}
where the linear operator $\hat F_W$ is specified in the appendix. The
spectrum of $\hat F_W$ is given by
$$
\lambda^{(W)}_m = \left\{ \begin{array}{ll} 0, &1\leq |m| \leq |W|,\\
    |m| - |W|,\enspace &|W|<|m|. \end{array}\right.
$$
To find $Z_W$, one has to integrate over the massive modes with
eigenvalues $\lambda^{(W)}_m$ as well as over the zero modes $z_a$.
The massive mode integration leads to a reduction of
the instanton action,
\begin{eqnarray}
S_{\rm inst}=\cG_0|W|\enspace\longrightarrow\enspace(\cG_0-\ln\frac{gE_c}T)|W|.
\end{eqnarray}
This ``renormalization'' of the
coefficient ${\cG_0}$ is analogous to the renormalization of the
conductance in the Ohmic model discussed in Sec.~\ref{subsec-phase}
(see Eq.~(\ref{eq-efetov}).) In the present context it signals that
instantons  become increasingly important at low temperatures.

Finally, the integration over  zero modes obtains a prefactor $\sim
(g\ln E_c/T)^{|W|}$. Combining all contributions and accounting for
combinatorial factors we thus obtain
\begin{eqnarray}
 \!\!\!\frac{Z_W}{Z_0}\!\! \eq\!\! \sum_{s=0}^\infty
\frac{1}{(s+|W|)!\,s!}\left(\pi g^2\frac{E_c}Te^{-{\cG_0}}
\ln\frac{E_c}T\right)^{2s+|W|} \!\!\! \!\!\! \!\!\!.\enspace
\end{eqnarray}
We finally sum over  winding numbers $W$ to obtain the instanton
contribution to the free energy~\cite{Grabert96}, $F=-T\ln Z$,
\begin{eqnarray}
\label{eq-freephase}
\delta F(q,T)=-2\pi g^2E_ce^{-\cG_0}\ln\frac{E_c}T\cos(2\pi q).
\end{eqnarray}
Charge quantization renders the free energy a periodic function of the
gate voltage $q$.  However, as expected, the amplitude is
exponentially small in $\cG_0$.

Using the same formalism, we may compute the conductance $G(q)$
through the dot. As shown in App.~\ref{app-conductance}, the phase
representation of the Kubo conductance is given by
\begin{widetext}
\begin{eqnarray}
G(q) =-2g^2Z^{-1}\sum_W e^{2\pi iWq}
\lim_{\omega\to0}\frac1\omega\Im\left[\left\langle|\langle\delta\phi|\hat
F_W|m\rangle|^2\right\rangle_{S_W[\delta\phi]}\right]_{i\omega_m\to\omega+i0}.
\end{eqnarray}
\end{widetext}
Summing over instanton
configurations, one obtains
\begin{eqnarray}
\label{eq-0d_cond} G(q)= \frac g2\left(1-\frac{\pi^3}3\frac{\tilde
E_c }T\cos(2\pi q)\right),
\end{eqnarray}
where $\tilde E_c=g^2E_c\exp[-\cG_0]$ may be interpreted as an
effective charging energy~\cite{Golubev96,Grabert96}. As with the free
energy, the conductance contains a weak gate voltage periodic
modulation.  Notice, however, that there is no zero--mode factor $\sim
\ln E_c/T$.  Rather, the massive mode integration leads to the much
stronger divergence $\tilde E_c/T$.

The above approach is valid as long as the correction is small,
i.e., $T\gg \tilde E_c|\cos(2\pi q)|$. At smaller temperatures,
the instanton expansion becomes uncontrolled. The $q$--dependence
of the corrections to the free energy and the conductance is
linked to charge quantization. In the following, we shall study
the same problem within the charge description. We will  show that
both models yield identical results, and point out a number of
similarities and differences between  the two approaches.

\subsection{Charge model}
\label{subsec-0d_charge}

Applied to a lead--dot--lead setup, the charge action
Eq.~(\ref{eq-matveev}) assumes the form
\begin{widetext}
\begin{eqnarray}
\label{eq-matveev1}
S\left[\theta\right] =
\eq\frac1T\sum_m \left(\sum_{i=L,R}\pi|\omega_m|\theta_{i,m}^{\;2} +
 E_{\rm c}(\theta_{L,m}-\theta_{R,m}-q \,\delta_{m,0})^2
\right) - \frac{D}\pi\sum_{i=L,R}r_i\int\limits_0^\beta d\tau\;
\cos(2\pi\theta_{i}(\tau)).
\end{eqnarray}
\end{widetext}
Generalizing to multi--channel  contacts, we replace
$r_i\to\exp[-\cG_{i,0}/2]$. As long as $\cG_{i,0}\gg1$, one may
treat the cosine--term perturbatively. Expanding to lowest
non--vanishing order in $e^{-\cG_{L,0}/2}$ and $e^{-\cG_{R,0}/2}$,
one obtains the contribution to the partition function
\begin{eqnarray}
\frac{Z_1}{Z_0}=\frac{D^2}{2\pi^2}e^{-\cG_0}\cos(2\pi q)\int\!
d\tau\,d\tau'\;\Re\left\langle
  e^{2i\pi(\theta_L(\tau)-\theta_R(\tau'))}\right\rangle,\nn 
\end{eqnarray}
where $\cG_0=(\cG_{L,0}+\cG_{R,0})/2$.

Using that $\langle e^{i\hat X}\rangle=\exp[-\frac12\langle\hat
X^2\rangle]$ and evaluating the correlators
$\langle\theta_i\theta_j\rangle$, we thus obtain
\begin{eqnarray}
\frac{Z_1}{Z_0}&\sim&E_ce^{-\cG_0}\cos(2\pi q)\int\!
d\tau\,d\tau'\;\frac1{|\tau-\tau'|}\\ 
\eq\frac{E_c}Te^{-\cG_0}\ln\frac{E_c}T\cos(2\pi q).
\end{eqnarray}
This correction corresponds to  the one--instanton correction to
the phase model. In a similar manner, the inclusion of higher
order terms yields the correction to the free energy
\cite{matveev},
\begin{eqnarray}
\delta F(q,T) \eq-\frac{8e^{\bf
    C}E_c}{\pi^3}e^{-\cG_0}\ln\frac{E_c}T\cos(2\pi q), 
\end{eqnarray}
where ${\bf C}\approx 0.577$ is the Euler constant.

The phase--dependent contribution $\delta F$ matches the result
obtained from the phase model, Eq.~(\ref{eq-freephase}).  Again,
however, the approximations leading to this result are limited to
temperatures higher than a certain cutoff temperature. Presently,
however, we are in a position to explore what happens below that
temperature:

At smaller temperatures, the cosine--potential itself provides a
mass to the field $\theta$. The relevant scale may be extracted by
comparing the amplitude of the cosine--potential with the
effective bandwidth --- they become comparable at $T\sim \tilde
E_c(q) = E_ce^{-\cG_0}|\cos(2\pi q)|$. This temperature scale
provides a cut--off for the logarithmic temperature--dependence of
the free energy~\cite{matveev}.  Thus, at temperatures $T<\tilde
E_c(q)$, the correction to the free energy saturates at $\delta
F(q,\tilde E_c(q))$.

Corrections to the conductance may be found in a similar way. As a
result, one obtains~\cite{matveev,ABG}
\begin{eqnarray}
G\eq \frac g2\left(1-\frac{e^{\bf C}E_c}{\pi T}e^{-\cG_0}\cos(2\pi q)\right).
\end{eqnarray}
To exponential accuracy, and using $\tilde E_c \sim E_ce^{-\cG_0}$,
this is identical to the result (\ref{eq-0d_cond}) obtained from the
phase model.

\subsection{Comparison}
\label{subsec-compare}

In the previous two sections, we reviewed Coulomb blockade effects
for a single quantum dot strongly coupled to the two leads.

In Sec.~\ref{subsec-0d_phase}, we started from a microscopically
derived  phase action. We applied an instanton analysis to
identify  the `effective charging energy' $\tilde E_c=g^2
E_c\exp[-\cG_0]$. We also found that  perturbation theory is
applicable for temperatures $T>\tilde E_c |\cos(2\pi q)|$.

In Sec.~\ref{subsec-0d_charge}, the starting point was a
phenomenological model for the same system. The derivation of that
model required the condition of weak backscattering in all
channels, $r_\alpha\ll 1$. To exponential accuracy, a perturbative
expansion in the reflection coefficients obtained results
equivalent to those of Sec.~\ref{subsec-0d_phase}. Again
perturbation theory breaks down when the effective amplitude of
the cosine--potential becomes of order of the bandwidth, $T\sim
\tilde E_c|\cos(2\pi q)|$. At lower tempertures, the temperature
dependence of the free energy saturates. The conductance, on the
other hand, is suppressed and behaves
as~\cite{averin-nazarov,matveev} $(T/\tilde E_c(q))^2$.

As we shall demonstrate below, the equivalence of the phase and
the charge description, respectively, will pertain to array--type
geometries. However, as with the single quantum dot, the charge
model will be preferable over the phase model when it comes to
exploring low--temperature properties. Remarkably, for arrays the
equivalence of the two descriptions may be demonstrated in
explicit terms (and not just exemplified on specific observables
as was the case for the single dot.) A second difference to the
isolated dot  regards the role of the cosine--potential of the
charge model: While for a dot, a strong potential (of the order of
the `bandwidth') is required to impede charge fluctuations, in an
array arbitrarily weak periodic potentials suffice to `pin'
spatial modulations~\cite{AGK,MKG}. In fact, the efficiency of
even weak potentials in impeding charge fluctuations will be at
the root of the localization phenomenon in arrays.


\section{1{{d}} array}
\label{sec-array-1d}

We next advance to the 1{{d}} array geometry. As in the previous section,
we start with the AES phase model (\ref{sec-1d_phase}) and then
continue to the charge model (\ref{sec-1d_charge}).

\subsection{Phase model}
\label{sec-1d_phase}

Our starting point is the generalization of  Eqs.~(\ref{eq-S_c}) and
(\ref{eq-AES}) to the 1{{d}} array geometry,
\begin{eqnarray}
S_c[\phi]\eq\sum_{j=1}^M\int d\tau
\,\left(\frac{\dot\phi_j^2}{4E_c}-iq\dot\phi_j\right),\\
S_t[\phi]\eq
\frac1{16}\sum_{j=1}^{M-1} \sum_k\kappa_k
\tr\left[\left(\Lambda
    e^{i\delta\phi_j}\Lambda e^{-i\delta\phi_j}\right)^k\right],
\end{eqnarray}
where $M$ is the number of grains, and
$\delta\phi_j=\phi_{j+1}-\phi_j$.

As in the single dot case, we shall focus on instanton phase
configurations.  Consider a configuration, where the phase on grain
$j_0$ has an instanton, i.e., a winding number $W_{j_0}=1$, whereas
$W_j=0$ everywhere else. The action of this configuration is given by
$S_t=2(\cG_0/2)$ from the tunneling terms
$j_0\to j_0\pm 1$ and $S_c=\pi^2T/E_c$ from the charging
term. Importantly, the fact that the tunneling action depends only on
phase {\it differences} implies the existence of other instanton
configurations of similar action: for example, the action of a
sequence of $W=1$ instantons on the
grains $\{j_0,j_0+L\}$ will again cost tunneling action $S_t=2(\cG_0/2)$
(from the tunneling terms
$j_0-1\to j_0$ and $j_0+L\to j_0+L +1$.) Only the charging
action depends in an extensive manner on  the length $L$ of the plateau:
$S_c(L)=L\pi^2T/E_c$. However, $\pi^2T/E_c\ll1$ whereas
$\cG_0\gg1$.

\begin{figure}[h]
  \centerline{\epsfxsize=3.25in\epsfbox{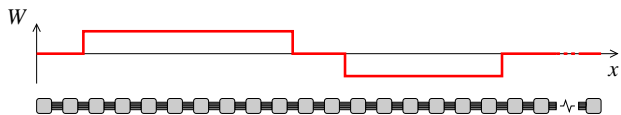}}
\caption{(Color online) A typical configuration of the instanton winding numbers
along the 1{{d}} array. The edges of the plateaus play the role of
fictitious charges in the Coulomb gas mapping.} \label{fig2}
\end{figure}

Thus, typical  instanton configurations consist of long plateaus
with a given winding number $W$ that differs from the winding
number of the background, $W_{\rm b}$, by $\pm1$. (Ignoring the
charging term, the action of a single plateau with height
$|W-W_{\rm b}|=2$ equals that of two $|W-W_{\rm b}|=1$ plateaus.
However, the latter configuration has the benefit of a much larger
entropy which is why we will ignore the unlikely formation of
configurations with step heights $|W-W_{\rm b}|>1$ throughout. The
action of the phase model for a given configuration of winding
numbers reads
\begin{eqnarray}
S[W_j]=\sum_j\left(\frac{\pi^2T}{E_{\rm c}} W_j^2 +
  {\cG_0\over 2}|W_j-W_{j-1}|\right).
\end{eqnarray}
The partition function is obtained by summing over all
configurations $\{W_l\}$,
\begin{eqnarray}
Z\eq\sum_{\{W_l\}}e^{-S\left[\{W_l\}\right]}\times({\rm
  fluctuation\enspace terms}). 
\end{eqnarray}
One may rearrange the sum by  introducing a  new set of variables
$\sigma_i=W_i-W_{i-1}\in\{-1,0,1\}$.
Boundary conditions at the leads require $W_0=W_{M+1}=0$, i.e.,
$\sigma_i$ obeys the sum rule
$\sum_{i=1}^{M+1}\sigma_i=0$. The contributions to the partition
function from configurations 
$\{\sigma_l\}$ can be classified according to $\sum_i\sigma_i^2=2k$
(where the sum is even due to the neutrality condition
$\sum_i\sigma_i=0$.) We thus find
\begin{eqnarray}
\label{eq-CG00}
Z\eq\sum_{\{\sigma_l\}}e^{\frac{\pi^2T}{E_{\rm
      c}}\sum_{j,j'}|j-j'|\sigma_j\sigma_{j'} -
  \frac12\cG_0\sum_j\sigma_j^2}\\
\eq\sum_{k=1}^\infty \left(e^{-\cG_0/2}\right)^{2k}
\!\!\!\!\!\!\sum_{\{\{\sigma_l\}|\sum_i\sigma_i^2=2k\}}
\!\!\!\!\!\! e^{\frac{\pi^2T}{E_{\rm
      c}}\sum_{j,j'}|j-j'|\sigma_j\sigma_{j'}},\nn
\end{eqnarray}
where we used that $|W_i-W_{i-1}|\to\sigma_i^2$ as well as
\begin{eqnarray}
\label{wint}
\sum_i W_i^2\eq-\sum_i\sum_{j\leq
    i}\sigma_j\sum_{j'>i}\sigma_{j'} = -\frac12 \sum_{j,j'}
  |j-j'|\sigma_j\sigma_{j'}.\nn
\end{eqnarray}
Since the action for $\sigma_i=0$ is zero, it is sufficient to focus
on non--zero charges $\sigma_i=\pm1$ at positions $x_i$. Relabeling
the $\sigma_i$, the summation over different configurations is
replaced by an integration over the positions $x_i$ of the charges
$\sigma_i$.
  The
corresponding partition function can be rewritten as
\begin{widetext}
\begin{eqnarray}
\label{eq-CG0}
Z\eq\sum_{k=1}^\infty\left(e^{-\cG(T)/2}\right)^{2k}
\frac1{(k!)^2}\prod_{i=1}^{2k}\,\sum_{x_i=1}^M\;
\exp\left[\frac{\pi^2T}{E_{\rm
    c}}\sum_{j,j'=1}^{2k}(-1)^{j+j'}|x_j-x_{j'}|\,
    \right],
\end{eqnarray}
\end{widetext}
i.e., as the  partition function of a classical 1{{d}} neutral
Coulomb gas  with the fugacity $e^{-\cG(T)/2}$ and effective
charge $\pi{T}/\sqrt{E_{\rm c}}\,$. Here we have replaced
$\cG_0\to \cG(T)$, anticipating that the integration over
fluctuations around the stationary instanton  configurations will
lead to a temperature--dependent renormalization of the
conductance. As  a result of a calculation conceptually similar to
that for the isolated dot geometry
(cf.~App.~\ref{sec:fluct-determ-1d}) we indeed find that the
fluctuation determinant is given by
\begin{eqnarray}
\label{F_1D}
{\cal F}\eq\exp\left[-\frac{2M}{M+1}\ln\frac{gE_{\rm
      c}}T\right]=\left(\frac T{gE_{\rm c}}\right)^{\frac{2M}{M+1}}.
\end{eqnarray}
The presence of this factor leads to the renormalization
$\cG_0\to{\cG(T)}=\cG_0-\frac{2M}{M+1}\ln(gE_{\rm c}/T)$ implied in
Eq.~(\ref{eq-CG0}). Notice that for a long array $M\to \infty$ the
renormalization is more significant than for a single grain, $M=1$.

It is well known (see App.~\ref{app-SG}) that the 1{{d}}
Coulomb gas, Eq.~(\ref{eq-CG0}), is equivalently described by the
action of the 1{{d}} discrete sine--Gordon model
\begin{eqnarray}
\label{eq-SG1}
S[\theta]=\sum_j\left\{\frac
{E_c}T(\theta_{j+1}\!-\!\theta_j)^2-2e^{-\cG(T)/2}\cos(2\pi
\theta_j)\right\}\!.\enspace 
\end{eqnarray}
Introducing the parameter
\begin{eqnarray}
\gamma=4\pi^2 g e^{-\cG_0/2}\ll 1,
\end{eqnarray}
and using that $e^{-\cG(T)/2}=(gE_c/T)e^{-\cG_0/2}$ this action may be
reformulated as
\begin{eqnarray}
\label{eq-CGres}
S[\theta]=\frac
{E_c}T\sum_j\left\{(\theta_{j+1}-\theta_j)^2-\frac\gamma{2\pi^2}\cos(2\pi
  \theta_j)\right\}. 
\end{eqnarray}
Before discussing the physics of this model, we will derive it in an
alternative manner, viz.~from the charge model. This will
make the equivalence of the charge and the phase description explicit,
and elucidate the physical meaning of the field $\theta$.

\subsection{Charge model}
\label{sec-1d_charge}

The straightforward generaliztion of the charge action
Eq.~(\ref{eq-matveev}) to a 1{{d}} array of dots is given by
\begin{widetext}
\begin{equation}
\label{eq-matveev1d}
S\left[\theta\right] = \sum_j \left\{\frac1T\sum_m
  \left( E_{\rm
      c}(\theta_{j+1,m}-\theta_{j,m}-q\,\delta_{m,0})^2
+ \pi|\omega_m|\theta_{j,m}^{\;2}\right) -
\frac{Dr}\pi\int d\tau\,\cos(2\pi\theta_j)\right\}.
\end{equation}
\end{widetext}
Thermodynamic properties of the array may be probed by
differentiation with respect to the gate voltage $q$. Noting that
$q$ couples only to the static
sector of the field $\theta_{j,0}=\int d\tau \;\theta_j(\tau)$, one may
try integrate out all non--zero Matsubara components
$\theta_{j,m\neq 0}$ at an early stage of the analysis. To this end let us
write
\begin{eqnarray}
\theta_j(\tau) = \theta_{j,0} + \delta\theta_j(\tau).
\end{eqnarray}
In the quadratic part of the action $\theta_0$ and $\delta\theta$
decouple. Denoting the average over the quadratic
$\delta\theta$--action by $\langle\dots\rangle$, we approximate the
functional integral over the anharmonic part of the action by
$\langle \exp(\int \cos\theta\rangle \rangle \simeq \exp \int \langle
\cos\theta \rangle$, i.e.,
\begin{eqnarray}
\langle\cos(2\pi\theta)\rangle_{m\neq0} &\to& {1\over 2} \left(
  e^{i2\pi\theta_0} \left\langle
    e^{i2\pi\delta\theta}\right\rangle + e^{-i2\pi\theta_0}
  \left\langle
    e^{-i2\pi\delta\theta}\right\rangle\right)\nn\\
&& = \cos(2\pi\theta_0) \exp\left[ -2\pi^2 \langle
  \delta\theta^2\rangle\right].
\end{eqnarray}
Performing the integral over $\delta\theta$, we find
\begin{eqnarray}
\label{int_mass_modes}
\langle\delta\theta^2\rangle=\frac
T{2M}\!\!\sum\limits_{|\omega_m|\neq 0}^D\!\sum\limits_
p\frac1{E(p)\!+\!\pi|\omega_m|}\!=\!\frac1{2\pi^2}\ln\frac
D{e^{\bf C}E_c},
\end{eqnarray}
where $E(p)=4E_c\sin^2(p/2)$ is the lattice dispersion.
Importantly, Eq.~\eqref{int_mass_modes} does not contain
temperature--dependent   
infra--red singularities. This provides the a posteriori
justification of the above integration procedure. Substituting
the $\delta\theta$--averaged cosine operator back into the action of
the static field we obtain
\begin{eqnarray}
\label{eq-CG_charge}
S_{\rm cl}={E_c\over T}\sum_{j=1}^{M-1} \left\{
  (\theta_{j+1,0}\!-\!\theta_{j,0}\!-\!q)^2 \!-\!\frac{\gamma}{2\pi^2}
  \cos(2\pi\theta_{j,0})\right\}\!\!,\enspace
\end{eqnarray}
where $\gamma=2\pi e^{\bf C}r$. As shown in appendix \ref{app-gamma},
generalization to the multi--channel case amounts to the substitution
$\gamma=2\pi e^{\bf C}\prod_\alpha r_\alpha\sim e^{-\cG_0/2}$.

To exponential accuracy, this result is equivalent to the action
Eq.~(\ref{eq-CGres}) obtained from the Coulomb gas representation of
the phase model.

\subsection{Thermodynamics of 1{{d}} arrays}

In the previous sections, we have shown that the long--range physics of
the granular array is effectively described by a classical model with
free energy $F=TS_{\rm cl}$, where
\begin{eqnarray}
\label{eq-FK}
F(q)=E_c\sum_j\left\{(\theta_{j+1}-\theta_j-q)^2 - \frac
  \gamma{2\pi^2}\cos(2\pi\theta_j)\right\}\!,
\end{eqnarray}
and the field index `0' has been dropped for convenience.

\begin{figure}[h]
  \centerline{\epsfxsize=3.25in\epsfbox{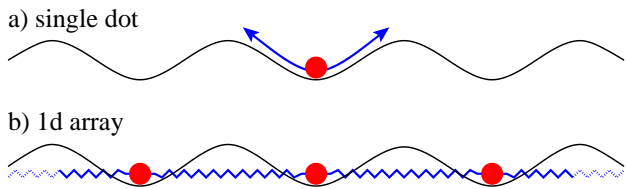}}
  \caption{(Color online) Chain of particles (each representing one quantum dot)
    subject to a shallow cosine--potential. While an isolated particle
    would experience no more than a weak
    back--scattering effect, a particle chain
    is subject to strong pinning. This leads to the
    appearance of a new (pinning) temperature scale $T^*\gg T_0$ which
    does not exist for single dots.} \label{fig3}
\end{figure}

This model is known as the discrete sine--Gordon model or, in the
context of the adsorption of atoms on a periodic substrate, as the
Frenkel--Kontorova model~\cite{Chaikin}. [The Frenkel--Kontorova model
describes a harmonic elastic chain of ``atoms" with stiffness $E_c$,
placed on top of a periodic ``substrate" potential of amplitude
$\gamma E_c/(2\pi^2)$. See Fig.~\ref{fig3}.] In the following, we
will summarize the physical properties of this model, and translate to
the context of the 1{{d}} quantum dot array.

In the absence of a gate voltage ($q=0$), the ground state of the
chain is described by $\theta_j=0$ for all $j$. This field describes a
zero charge configuration.  Charge excitations correspond to
non--vanishing solutions extremizing the free energy $F$. Varying the
action one finds that these configurations are given by
\begin{eqnarray}
\label{eq-sol1} \bar\theta_j = \frac2\pi\, {\rm
arctan}\left[\exp\left\{\sqrt
    \gamma (j-j_0)\right\}\right].
\end{eqnarray}
Eq. \eqref{eq-sol1} describes a solitary excitation centered at
coordinate $j_0$ and extended over a scale $\xi_{\rm s}=1/\sqrt
\gamma\gg1$. The total charge carried by this excitation is quantized
and given by $\cN_{\rm
  tot}= \sum_j (\theta_{j+1}-\theta_j)=\theta_{M+1}-\theta_1=1$. (Due to
the large inter--grain tunneling,) it is spread over a large number
$\gamma^{-1/2}\sim e^{\cG_0/4}$ of grains.

Substituting this solution into the free energy (\ref{eq-FK}), one
obtains the soliton energy
\begin{equation}
\label{Tstar}
T^* = E_c\sqrt \gamma\gg T_0\, .
\end{equation}
At finite temperatures, the system will host a gas of thermally
excited solitons and anti-solitons with density $n_{\rm
  s}\sim\exp[-T^*/T]$. Due to the absence of gapless charge carriers
in the system, transport will exhibit activation behavior, as we will
show below.

At finite gate voltage, the uniform configuration $\theta_j=0$ will
acquire the finite energy $\delta F_{\rm el}(q)=ME_cq^2$ (the elastic
term in the action.) By contrast, the configuration $\theta_j=jq$
which minimizes the elastic term has the energy $\delta F_{\rm
  pot}(q)=-ME_c\gamma/(2\pi^2)$ (the cosine--potential.) The smaller
of the two determines the ground state of the system. In the language
of the Frenkel--Kontorova model, the ``incommensurability parameter"
$q$ represents the periodicity mismatch between the chain and the
substrate.  For small values of $q$ the system will find it favorable
to retain a commensurate state, i.e., the chain will stretch a little
so as to still benefit from an optimal coupling to the substrate.
Thus, the system remains in a commensurate phase $\theta_j=0$ for gate
voltages $q<q^*=\sqrt{\gamma/(2\pi^2)}$. At $q^*$, where the two
energies become equal ($\delta F_{\rm el}(q^*)=\delta F_{\rm
  pot}(q^*)$), a commensurate--incommensurate transition takes place
--- at this gate voltage solitons are created at no cost. For the
average number of electrons per grain, $\bar \cN(q) \equiv
q-\partial_q F/(2ME_c)$, one thus expects: $\bar \cN(q) =0$ for
$|q|\leq q^\ast$ (insulator) and $ \bar \cN(q) \to q$ for $|q|>q^\ast$
(metal).

\begin{figure}[h]
  \centerline{\epsfxsize=2.25in\epsfbox{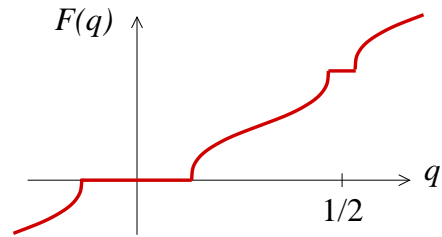}}
\caption{(Color online) Free energy of the 1{{d}} array as function of the common
background charge, $q$. Only two insulating plateaus at $q=0$ and
$q=1/2$ are shown.} \label{fig4}
\end{figure}

A more thorough discussion of the system
(cf.~Ref.~[\onlinecite{Chaikin}]) shows that insulating `plateaus'
along with superimposed solitary excitations form not only around
$q=0$, but also around other rational values of $q$.  However, both
the width of these plateaus and the corresponding activation energies
decrease for higher rational fractions. Among the low--lying rationals,
$q=1/2$ plays a particularly interesting role. Indeed, for a {\em
  single} grain, $q= 1/2$ represents the charge degeneracy point, where
the system is in a conducting state (Coulomb blockade peak).
Unexpectedly, the {\it array} exhibits a very different behavior.
Using the language of the Frenkel--Kontorova model, $q=\pm 1/2$ is
special in that the 
atoms of the unperturbed chain alternatingly find themselves in
minima/maxima of the substrate potential.  Under these conditions,
energy can be gained by a `Peierls--distortion', doubling the
period, i.e., a density modulation with amplitude $\delta
\theta_j\sim \gamma$. This configuration does not respond to small
variations in $q$ --- corresponding to insulating behavior.
However, the width of the insulating plateau $\Delta q_{1/2} \sim
\gamma$ is much smaller than the width of the plateau $ \Delta
q_{0}\equiv 2q^\ast\sim \sqrt{\gamma}$ at $q=0$. The dependence of the
free energy on $q$ (including the two plateaus) is shown in Fig.~\ref{fig4}.

\subsection{DC transport}

In order to discuss the DC conductivity of the array, one needs to
restore the low--frequency, $\omega\ll T$, dynamics of the
classical charge model, Eq.~(\ref{eq-FK}). In principle, this may
be done by keeping the dissipative term in the action.  It turns
out that in the multi--channel case (cf. ~App.~\ref{app-gamma}),
and for strong backscattering, the coefficient of the dissipative
term reads as $\pi g^{-1} |\omega_m|\theta_{j,m}^{\;2}$. While
this seems to satisfactorily describe the dynamics of Ohmic
dissipation, the main drawback of the imaginary time approach is
that it involves cumbersome analytical continuations $\omega_m \to
\omega+i0$. To avoid this complication, it is convenient to pass
to the Keldysh representation. In fact, it turns out to be
sufficient to focus on the semi--classical limit of the Keldysh
formalism, i.e., a limit physically equivalent to a description of
the system in terms of a classical Langevin
equation~\cite{Kamenev01}.  Referring for a more detailed
discussion of these connections to App.~\ref{app-Keldysh}, we here
take the adequacy of this formulation for granted and describe the
system in terms of the Langevin equation right away:

To start with, consider the equations of motion of the static model,
Eq.~(\ref{eq-FK}): $\partial\left({e\over 
    C}\partial\theta_j\right)-{e\gamma\over 2\pi C}\sin
(2\pi\theta_j)=0$.  Since $\frac eC\partial\theta_j\equiv V_j$ is the
voltage on grain $j$, this equation simply expresses the fact that in
the absence of charge quantization, $\gamma\to 0$, all grains are
equipotential: $\partial V_j=V_{j+1}-V_j=0$. We now ask how these
equations have to be modified if currents
are allowed to flow. The minimal classical description
of dissipative current flow between grains $j+1$ and $j$ is provided by
the current--voltage relation
$V_{j+1}-V_j=RI_j$, where $R=2\pi \hbar /(e^2 g)$
is the contact resistance, and $I_j=e\partial_t \theta_j$  the
current flowing between grains $j$ and $j+1$. Restoring the
$\gamma$--term, we are  led to the phenomenological generalization of
the equation above, 
\begin{eqnarray}
                                                \label{eq-langevin}
\frac{\pi}{g}\, \partial_t\theta -
E_c\Big[\partial^2\theta - \frac{\gamma}{2\pi}
\sin(2\pi\theta)\Big]
 = -\frac{e}{2}\, E +\xi(t),
\end{eqnarray}
where we have introduced a Gaussian noise term, $\xi(t)$, with the
correlator
\begin{equation}
                                            \label{eq-noise}
\langle \xi_j(t)\xi_{j'}(t')\rangle  = \frac{2\pi
  T}{g}\, \delta(t-t')\delta_{j,j'}\, ,
\end{equation}
to satisfy the fluctuation--dissipation theorem, and an external
electric field, $E$, as a driving source of current. Formally,
Eq.~\eqref{eq-langevin} represents a Langevin equation for the
classical degree of freedom, $\theta$.

Our goal is to calculate the current, $I$, driven by  a weak
uniform field, $E$. As we saw above, charge transport in the present
model is by solitary excitations. As an ansatz for the current we thus
use $I=en_{\rm s} v$, where $n_{\rm s}$ is the soliton
concentration and $v$ is their effective drift velocity. While the
soliton concentration, $n_{\rm s}$, has been discussed above, the
drift velocity still needs to be determined. To this end, we
temporarily ignore the noise term and seek for
propagating solutions
of Eq.~(\ref{eq-langevin}). Assuming the external field to be weak, we
consider the ansatz
$\theta(j,t)=\bar\theta(j-vt)+\theta_1(j-vt)+\zeta$, where $\bar
\theta$ is the static soliton, $\theta_1\sim E$  a small
distortion of the soliton shape due to the presence of the
external field, and the constant $\zeta$ accounts for the weak
shift of the minimum of the periodic potential by the field:
$E_c\gamma\sin(2\pi\zeta)=\pi E$.  Linearizing
Eq.~(\ref{eq-langevin}), we find that $\theta_1$ satisfies the
equation
\begin{equation}
                               \label{eq-linearized}
E_c \hat {\cal L}_{\{\bar\theta\}}\theta_1=\frac
  v{g}\partial\bar\theta -\frac
  E{\pi}\sin^2(\pi\bar\theta),
\end{equation}
where $\hat {\cal L}_{\{\bar\theta\}} \equiv
\partial^2 -\gamma \cos(2\pi\bar\theta)$. Importantly, it is not our prime
objective to identify the actual shape of the soliton
($\theta_1$); Rather, we whish to compute its speed of propagation,
$v$. An equation for $v$ may be obtained by noting that
$\theta_1$ has been introduced  to describe a change in the {\it
  shape} of the soliton 
in response to the field. This needs to be distinguished from the
temporal change in its {\it position} (which has been accounted for by
the introduction of the as yet undetermined shift $vt$.) To
discriminate between these two effects, we require that the linear
equation determining $\theta_1$ be an equation in a function space
orthogonal to the zero mode function, $\partial \bar \theta$, describing
translations of the soliton, $\bar \theta(x+\delta x) -\bar
\theta(x)\sim \delta x \partial \bar \theta$. In
particular, $v$ should be determined in such a way that the r.h.s. of
the equation be orthogonal to $\partial\bar \theta$,
\begin{equation}
\int dx\left(\frac
  v{g}\partial\bar\theta -\frac
  E{\pi}\sin^2(\pi\bar\theta)\right)\partial\bar\theta=0.
\end{equation}
Using Eq.~(\ref{eq-sol1}) for $\bar\theta$, this requirement leads to
the expected result $v\sim gE$. This in turn implies that the  DC
conductivity is given by 
$\sigma=gn_{\rm s}(T)$. Using that for low temperatures, $T<T^\ast$,
the soliton density shows activation behavior, $n_{\rm s}(T)\sim
\exp[-T^*/T]$, we arrive at the result Eq.~(\ref{eq-res}).

\subsection{Density of states}

As another quantity of interest we discuss the (tunneling) density of
states.
At large energies, the density of states can be conveniently described
within the phase model. Within the framework of that model, the
DoS $\nu(\epsilon) = -{1\over \pi}\Im {\,\rm
  tr}\left[G(i\omega_m)\right]\big|_{i\omega_m \to \omega+i0}$ is represented as
\begin{eqnarray}
\label{eq-dos}
\nu(\epsilon)=\nu_0T \, \Im\!\int\!\! d\tau\frac{e^{i\epsilon_n\tau}}
{\sin\pi T\tau}\!\left. \left\langle
    e^{i\sum\limits_q(\phi_q(\tau)-\phi_q(0))}\right\rangle
\right|_{i\epsilon_n\to\epsilon^+}\!\!\!\!\!.
\end{eqnarray}
At large temperatures,  the `Debye--Waller factor' $\left\langle
  e^{i\sum_q(\phi_q(\tau)-\phi_q(0))}\right\rangle\simeq
\exp[-\frac12\sum_q\langle|\phi_q(\tau)-\phi_q(0)|^2\rangle]$ may be
computed from the quadratic approximation to the phase action. This
leads to
\begin{eqnarray}
\sum_q\langle|\phi_q(\tau)-\phi_q(0)|^2\rangle
\eq 4T\sum_{q,m} \frac{1-\cos\omega_m\tau}
  {|\omega_m|(\frac g\pi q^2+\frac{|\omega_m|}{E_c})}\nn\\
\eq2\sqrt{2\frac{E_c\tau}g}.
\end{eqnarray}
Integrating over $\tau$ and performing the analytic continuation we
thus obtain the result
\begin{eqnarray}
\nu(\epsilon)\sim\nu_0\,\exp\left[-\sqrt{2\frac{E_c}{g\epsilon}}\right].
\end{eqnarray}
Notice that the conductivity and the DoS, respectively, are
governed by different energy scales. The DoS becomes exponentially
small at energies $\epsilon\sim E_c/g\gg T^*$. To understand what
happens below that energy, notice that the minimal energy required
to add a charge to the array is the excitation energy of a
soliton. Therefore, at zero temperature, the DoS vanishes at
energies smaller than $T^*$; i.e. $\nu(\epsilon<T^*)=0$. The
charge quantization leads to a hard gap in the DoS.

This concludes our discussion of the one--dimensional array. We
have found that the proliferation of instantons at low
temperatures drives the system into an insulating phase where
transport is activated  and conducted  by the charge solitons. The
temperature at which activation behavior sets in is exponentially
small in the dimensionless conductance. Yet it is parametrically
larger than the scale $T_0$ where the perturbative corrections
become large: $T^*/T_0\sim e^{\cG_0/4}\gg 1$. The tunneling DoS is
significantly suppressed at even higher energies and displays a
hard gap at the soliton energy.


\section{2{{d}} arrays}
\label{sec-array-2d}

We next extend our discussion to two--dimensional quantum dot
arrays. Our strategy will parallel that of the previous sections:
Starting from the phase representation we will establish a connection
to the complementary charge representation, and then discuss the
physics of the system in terms of the latter.

\subsection{Phase model}

\begin{figure}[h]
  \centerline{\epsfxsize=2.75in\epsfbox{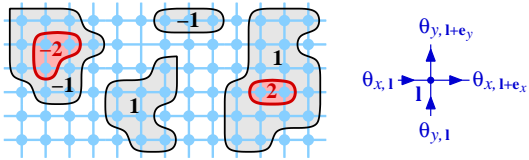}}
\caption{(Color online) The ``islands'' with fixed values of the winding number
in a 2{{d}} array. The area $A=1,2,\ldots $ is the number of grains inside
an island, while the circumference $L=4,6,8,\ldots $ is the number
of links crossed by its boundary.} \label{fig5}
\end{figure}

Again we start from the action (\ref{eq-S_c}) and (\ref{eq-AES}),
where the lattice summation now extends over a two--dimensional
regular array. Anticipating the importance of `winding numbers',
we begin with an identification of instanton solutions right away.
As depicted in Fig.~\ref{fig5}, instanton configurations in the
2{{d}} geometry assume the form of ``islands'', i.e., regions of a
certain winding number $W_{\rm island}$ surrounded by a background
with a different winding number $W_{\rm b}$. For entropic reasons,
configurations with $W_{\rm b}=0$ and $W_{\rm island}=\pm 1$ will
dominate. The action of one such island of area $A$ and
circumference $L$ is given by $S_{\rm
  island}=\frac{\pi^2T}{E_c}A+\frac {\cG_0}2L$. For an island
differing by a generic winding number from its background, this
generalizes to
\begin{eqnarray}
\label{eq-S_island}
S_{\rm island}=\frac{\pi^2T}{E_c}AW^2+\frac{ \cG(T)}2L|W|,
\end{eqnarray}
where we anticipated that the inclusion of fluctuations will again
manifest itself in a renormalization of the conductance, $\cG_0\to
\cG(T)$.

As a result of a straightforward
generalization of the one--dimensional fluctuation determinant
discussed in App.~\ref{sec:fluct-determ-1d}, we  indeed find
(cf.~App.~\ref{sec:fluct-determ-2d}) that the determinant due to
fluctuations around the stationary configuration is given by
\begin{eqnarray}
\label{Fdet_2D}
{\cal F}\eq\exp\left[-\frac L2 \ln\frac{gE_{\rm c}}T\right]=\left(\frac
    T{gE_{\rm c}}\right)^{\frac L2}.
\end{eqnarray}
We  account for the presence of this factor by renormalization
$\cG_0\to \cG(T)\equiv \cG_0-(2/{{d}})\ln(gE_c/T)$, where
${{d}}=2$. (In fact, it is straightforward to verify that this
result holds for systems of arbitrary dimensionality.)

\subsection{Charge model}
\label{sec-2d_charge}

Starting from Eq.~(\ref{eq-matveev})  (where the sum now extends
over a 2{{d}} array), our strategy will be to successively
integrate over high--frequency fluctuations to derive an effective
low--energy action. (Unlike in the 1{{d}} case where all dynamical
fluctuations could be integrated out in one sweep.) Consider,
thus, fluctuations of $\vec\theta(\tau)$ in a window of
(Matsubara) energies between the bandwidth $D$ and $\tilde D<D$.
Integration over these fluctuations will lead to a renormalization
of the cosine--potential. As long as the renormalized
backscattering amplitude is small, functional averages over high
frequency fluctuations can be taken using the quadratic action.

Due to the vector nature of $\vec\theta$, there are two types of
modes contributing to the fluctuations. Using the representation
$\vec\theta=\nabla{\chi}+\nabla\times\eta$, where $\chi$ and $\eta$
are scalar fields and $\nabla $ and $\nabla \times$ lattice variants
of gradient and curl, respectively, the charging action
takes the $\eta$--independent form $S_c[\chi,\eta] =E_c\sum_{\bf
  l}(\nabla \chi_{\bf 
l})^2/T$. The Gaussian average
\begin{eqnarray}
\langle\theta_i^2\rangle_{\rm f}\eq \frac1{2\pi}\sum_{\bf
  p}T\sum_{\tilde D<|\omega_m|<D}
\left(\frac1{|\omega_m|}+\frac\pi{E(\bp)+\pi|\omega_m|}\right), \nonumber
\end{eqnarray}
thus splits into two contributions, where only one (the
$\chi$--contribution) couples to the lattice dispersion
$E(\bp)=4E_{\rm c}(\sin^2\frac{p_x}2+\sin^2\frac{p_y}2)$.
 For frequencies $\tilde
D<E_{\rm c}$, we find
\begin{eqnarray}
\langle\theta_i^2\rangle_{\rm f}
\eq\frac1{4\pi^2}\ln\frac{D^2}{E_c\tilde D},
\end{eqnarray}
i.e., the $\chi$--mode is  effectively frozen out due to its coupling
to the charging action. However, (and unlike in the 1{{d}}--case) there is
one ``massless'' mode whose fluctuation amplitude depends on the
effective bandwidth 
$\tilde D$. Using this result to renormalize the
coefficient of the cosine--potential we obtain
\begin{eqnarray}
D\, e^{-\cG_0/2}\to \sqrt{E_c\tilde D}\,e^{-\cG_0/2}.
\end{eqnarray}
This integration procedure can be iterated until the renormalized
backscattering amplitude is of order of the bandwidth, i.e., $\tilde D\sim
E_ce^{-\cG_0}\equiv T_0$, corresponding to an effective
conductance $\cG_{\rm eff}={\cal O}(1)$. For temperatures
$T>T_0$, (i.e., for a Matsubara frequency spacing $\sim T$ larger than
the cutoff energy) one may approximate the action by the zero
Matsubara (classical) contribution
\begin{equation}
\label{eq-classical}
S_{\rm cl}[\vec\theta] = \frac{E_c}T \sum_{\bf l}
\Big\{(\nabla\cdot\vec\theta_{\bf l})^2 -
\frac{\gamma(T)}{2\pi^2}\sum_{i=x,y}\cos(2\pi\theta_{i,{\bf l}})\Big\},
\end{equation}
where $\gamma(T)=2\pi\sqrt{T/E_c}e^{-\cG_0/2}$.

At smaller  temperatures $T<T_0$, all modes become massive due to
the cosine--term. This leads to a saturation of the coefficient
$\gamma(T)$,
\begin{eqnarray}
\label{eq-gamma_ofT}
\gamma(T)\sim \begin{cases}\sqrt{T/E_c}\,e^{-\cG_0/2}&T>T_0,\\
\sqrt{T_0/E_c}\,e^{-\cG_0/2}\simeq e^{-\cG_0}&T<T_0.
\end{cases}
\end{eqnarray}
We shall show now that the perturbative expansion in powers of
$\gamma(T)$ of the classical model, Eq.~(\ref{eq-classical}), leads to
the same island scenario discussed above within the phase model.

For bookkeeping purposes, we label the coefficients
$\gamma_{\underline{\alpha}}(T)$ by the index of the bond
$\underline{\alpha}=(i,{\bf l})$ ($i=x,y$) to which they belong.  The
partition function can then be written as
\begin{eqnarray}
Z=\sum_{n=0}^\infty Z_n\, (E_c/T)^{n}\!\!\!\!
\prod_{\underline{\alpha}_j
  \atop(j=1\dots
  n)}\gamma_{\underline{\alpha}_j}(T),\nonumber
\end{eqnarray}
where $Z_n$ is a product of $n$ cosine--terms averaged over the
quadratic action $S_c[\vec \theta] =E_c\sum_{\bf l}(\nabla\cdot\vec
\theta_{\bf l})^2/T$.  Two types of contributions to this expansion
can be distinguished:
a) terms containing higher powers of the cosine taken at the same link
and b) terms involving different links. One may show that the first
class of terms describes perturbative corrections to the conductance
of a single contact (equivalent to those obtained in
Ref.~[\onlinecite{tschersich}] within the framework of the phase model.)
We here focus on the phenomenon of ``island
formation'', as described by the second family, b).

\begin{figure}[h]
  \centerline{\epsfxsize=1in\epsfbox{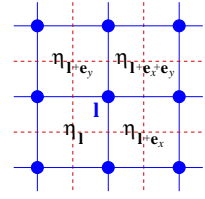}}
\caption{(Color online) The rotational field $\eta$. The lattice curl is given as
$(\nabla\times\eta)_{\bf l}
=\left({\eta_{{\bf l+e}_y}-\eta_{\bf l}\atop -\eta_{{\bf
        l+e}_x}+\eta_{\bf l}}\right)$. Going around one grain, the
$\eta$--fields   coming from the 4 
links cancel --- thus, yielding a non-vanishing contribution to
the expansion in $\gamma$.}\label{fig6}
\end{figure}

The expansion of
$Z_n$ generates expressions of the form $\exp[2\pi
i(\sum_{j_x=1}^s(\pm)\theta_{x,{\bf
    l}_{j_x}}+\sum_{j_y=s+1}^n(\pm)\theta_{y,{\bf l}_{j_y}})]$, where
$j_i$ labels contacts in $i$--direction. To perform the
$\theta$--average, we again use the representation
$\vec\theta=\nabla{\chi}+\nabla\times\eta$.  While the fields
$\chi$ are defined on the lattice, the fields $\eta$ are defined
on the reciprocal lattice (see Fig.~\ref{fig6}.) Importantly, the
(now classical) rotational field $\eta$ is strictly massless. This
means that contributions to the expansion containing an
uncompensated $\eta$--amplitude in the exponent average to zero.
The surviving terms obey the condition $\sum_{j_x}\pm(\eta_{\,{\bf
l}_{j_x}\!\!+{\bf
    e}_y}-\eta_{\,{\bf l}_{j_x}}) +\sum_{j_y}\mp(\eta_{\,{\bf
    l}_{j_y}\!\!+{\bf e}_x}-\eta_{\,{\bf l}_{j_y}})=0$. This
corresponds to the island structure: the lowest--order non--local term
is proportional to $\gamma^4=\gamma_{x,{\bf l}}\gamma_{x,{\bf
    l+e}_x}\gamma_{y,{\bf l}}\gamma_{y,{\bf l+e}_y}$ --- involving all
the four links surrounding grain ${\bf l}$.

Every island is weighted by a factor $(E_c\gamma(T)/T)^L$, where $L$
is the order in perturbation theory required for its formation. For an
island with winding number $|W|=1$, $L$ is the lattice
circumference. Islands with $|W|>1$ are surrounded by a chain of
$\gamma$--amplitudes $|W|$ times; in this case, $L=|W|\times$
circumference.
Finally, the averaging over the massive $\chi$--fields
results in a factor $\exp[-\pi^2TAW^2/E_c]$, where $A$ is the
island area.

Summarizing, every island carries a factor $\tilde P_W(A,L)\!=\!
\left(e^{-\pi^2T/E_c}\right)^{A W^2} \!\left(E_c\gamma(T)/ (2\pi^2
  T)\right)^{L |W|}$. Using Eq.~(\ref{eq-gamma_ofT}), one finds
that in the high--temperature regime, $T>T_0$,  $E_c\gamma(T)/
(2\pi^2 T)\simeq e^{-\cG(T)/2}$. Thus, $\tilde P_W(A,L)\!=\!
\exp\left[-\pi^2(T/E_c)A W^2-\frac12\cG(T)L |W|\right]$, which is
in perfect agreement with the prediction of the phase model,
cf.~Eq.~(\ref{eq-S_island}). At lower temperature, non--linear
fluctuation corrections in the phase model
diverge~\cite{tschersich} and the non--interacting instanton
treatment runs out of validity. However, having established the
equivalence of the phase and the charge model at $T>T_0$, one may
proceed with the analysis of the latter even at smaller
temperatures.

\subsection{Solitons and the BKT crossover}

In order to extract the low--temperature behavior of the array, we
investigate the properties of the classical model
(\ref{eq-classical}).  The lowest energy configuration of the action
is given by the homogeneous solution $\vec\theta=0\;(\mbox{mod}\;1)$
everywhere.  In order to minimize the cosine--potential, localized
excitations must have integer $\vec\theta$ far away from the core.
Since the total charge of such a localized excitation is
$e\!\int(d^2l)\,\nabla\cdot\vec\theta=e\!\int d\vec s\cdot\vec
\theta$, where the line integral on the r.h.s. is calculated over a
distant contour enclosing the excitation, the charge of the excitation
is quantized in integer multiples of $e$. Excitations of
lowest energy have charge $\pm e$. They consist of a large
(i.e., spread out over $\sim 1/\gamma\sim e^{\cG_0}\gg 1$ grains --- see below)
localized 2{{d}} soliton, connected to a 1{{d}} string of links with
$\theta_i=1$.  The other end of the string may either go to the system
boundary, or terminate in an anti-soliton of opposite charge. The
soliton solution centered at ${\bf l}=0$ can be written in the form
$\vec\theta_{\bf l}=1-\vec\vartheta({\bf l})$ for the links along the
string and $\vec\theta_{\bf l}=\vec\vartheta({\bf l})$ everywhere
else, where $|\vec\vartheta(|{\bf l}|\to\infty)|\to0$.  Minimization
of the action, Eq.~(\ref{eq-classical}), with respect to
$\vec\vartheta$ yields the saddle point equation for the soliton
solution,
\begin{equation}
\nabla(\nabla\cdot\vec\vartheta)-{\gamma(T)\over 2\pi} \sum_{i=x,y}\sin
(2\pi\vartheta_i){\bf e}_i=0.
\label{saddlepoint}
\end{equation}
Except for a domain consisting of ${\cal O}(1)$ links closest to the
core of the soliton, $\vartheta$ is small, justifying an expansion of
the sine--term in the saddle point equation, i.e.,
$\nabla(\nabla\cdot\vec\vartheta)-\gamma\vec\vartheta=0$. Its
unit--charge solution is
\begin{eqnarray}
                                     \label{eq-soliton}
\vec\vartheta({\bf l}) \eq -\frac{\sqrt\gamma}{2\pi}\,
K_1\!\left(\frac l{\xi_{\rm
  s}}\right)\,{\bf e}_l,
\end{eqnarray}
where ${\bf e}_l\equiv {\bf l}/l$ and $K_1$ is a modified Bessel
function. The large size of the soliton, $\xi_{\rm
  s}=1/\sqrt{\gamma(T)} \gg1$, justifies the continuum approximation.

To obtain the soliton energy $T^*$, we substitute this solution
back into the action, Eq.~(\ref{eq-classical}). One finds that the
energy originates primarily from the cosine--potential part and is
given by $T^*= (E_c\gamma(T)/(2\pi))\ln \xi_{\rm s}$. The large
factor $\ln \xi_{\rm s}=-{1\over 2} \ln\gamma(T)\gg 1$ is due to
the logarithmic spreading of the charge density over the wide
range of distances $1<l<\xi_{\rm s}$. Due to this factor $T^*\gg
T_0\sim E_c\gamma(T_0)$ (while their ratio is exponentially large in
$g$ in 1{{d}}, here 
it is only algebraic.) At larger distances, $l>\xi_{\rm s}$, the
charge density decays exponentially.

The charge spreading leads to a long--ranged soliton--soliton
interaction.  The solitons interact logarithmically up to a distance
$\xi_{\rm s}$ beyond which the interaction is exponentially screened.
Since the density of thermally--excited solitons is $n_{\rm s}\approx
\exp[-T^*/T]$, the mean distance between them is $l_{\rm
  s}=n_{\rm s}^{-1/2}\approx \exp[T^*/(2T)]$ which becomes
comparable to $\xi_{\rm s}$ at $T\approx T^*/(2\ln \xi_{\rm s})=
E_c\gamma(T)/(4\pi)$.  This condition is satisfied at temperatures
about the ``freezing'' temperature, $T\sim T_0$.  Thus, at $T < T_0$,
the thermally--excited charges are essentially non--interacting, while,
at $T>T_0$, there is a neutral (in average) gas of logarithmically
interacting solitons and anti-solitons. The soliton core energy yields
the fugacity $f$ of the logarithmic gas: $\ln f\simeq E_c\gamma(T)/T$.

In the latter regime the partition function of the charged degrees of
freedom may be written  as
\begin{equation}
                                       \label{eq-partition}
Z=\sum\limits_{n=0}^{\infty}\frac{f^{n}}{n!} \int (d^2l_1)\ldots
(d^2l_n)\,e^{\pm\frac{E_c\gamma(T)}{2\pi T}\sum\limits_{k,k'}^n
\ln|{\bf l}_k-{\bf l}_{k'}|}.
\end{equation}
The plus/minus signs in the exponent correspond to soliton--soliton
and soliton--anti-soliton interactions, respectively.

It is well known that the Coulomb gas in 2{{d}} described by
Eq.~(\ref{eq-partition}) undergoes the BKT
transition~\cite{BKT-b,BKT-kt} at a critical temperature $T_{\rm
  BKT}\approx E_c\gamma(T_{\rm BKT})/(4\pi)$.  For $T<T_{\rm BKT}$,
the charges are 
bound in charge--anti-charge pairs. In this regime, the finite
interaction range 
$\xi_{\rm s}$ leads to an exponentially small residual density of free
charges $n_{\rm s}\approx \exp[-T^*/T]$, where $T^*=T_{\rm
  BKT}\ln (\xi_{\rm s}^{\;2})$. Above the transition/crossover
temperature the density of free charges rapidly increases
as~\cite{BKT-kt} $n_{\rm s}\sim\exp[-2b\sqrt{T_{\rm BKT}/(T-T_{\rm
    BKT})}]$, where $b$ is a constant of order unity, driving the
array into the conducting phase.

Notice that the Coulomb interactions in our model are strictly
on--site (only the self--capacitance, $C$, is included.) The long
range of the soliton--soliton interactions is due to the fact that in
a strongly coupled array, $\cG_0\gg 1$, the charge is spread over a
large distance $\xi_{\rm s}\sim\exp[\cG_0/2]$.  To describe the truly
long--range Coulomb interactions, one may modify the model by
including mutual capacitances $C'$ between neighboring grains. It is
straightforward to show that such modification alters the range of
logarithmic interactions as $\xi_{\rm s} \to \xi_{\rm
  s}\sqrt{1+C'/C}$, while the charging energy now reads
$E_c=e^2/(2(C+C'))$.  In the limit $C\to 0$, while $C'$ remains
finite, the interaction range diverges, $\xi_{\rm s}\to \infty$, i.e.,
one deals with the true (logarithmic) 2{{d}} Coulomb interaction (without
the self--capacitance no electric field lines can leave the
system.) In this case, one observes a genuine BKT phase transition:
$\xi_{\rm s}\to \infty$ and the density of free charges below $T_{\rm
  BKT}$ becomes strictly zero.

\begin{figure}[h]
  \centerline{\epsfxsize=2in\epsfbox{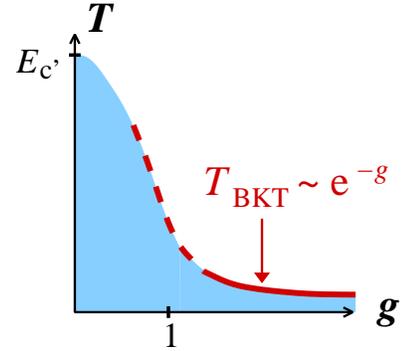}}
\caption{(Color online) Phase diagram of the 2{{d}} granular array. The thick line
indicates the BKT crossover temperature at $g>1$. At smaller $g$
the crossover disappears (in the self--capacitance model.) The
shaded area indicates the region with exponentially small
conductivity.} \label{fig7}
\end{figure}

The conductivity of the array may be evaluated in the same way as in
the 1{{d}} case. With the soliton velocity $v\sim gE$, one finds
$\sigma\simeq g\exp[-T^*/T]$ at temperatures below $T_{\rm BKT}$.
Above $T_{\rm BKT}$, the conductivity behaves~\cite{mooij} as $\sigma \simeq
g\exp[-2b\sqrt{T_{\rm BKT}/(T-T_{\rm BKT})}]$, whereas, at even
higher temperatures, this behavior crosses over to the
result~\cite{tschersich} of the perturbative calculation,
$\sigma=g-\ln(gE_c/T)$.

The corresponding phase diagram is shown in Fig.~\ref{fig7}.
Unlike previous works~\cite{fs2,fs,foot1} that predicted a
zero--temperature metal for $g>g_c\simeq 1$, we find that the
low--temperature phase is an insulator for arbitrarily large $g$.
The critical temperature, $T_{\rm
  BKT}(g)$, however, drops sharply at $g\simeq 1$ and, at large $g\gg
1$, behaves as $T_{\rm BKT}\sim E_cg \exp[-\cG_0]$.

\subsection{Finite gate voltage}

So far we have restricted ourselves to the case of zero gate voltage
only. A finite gate voltage induces a continuous background charge
$q\propto V_{\rm gate}$ on the grains. In this case the charging term
in the action Eq.~(\ref{eq-classical}) has to be replaced with $S_{\rm
  cl}^{\rm (c)}[\vec\theta;q]=(E_c/T) \sum_{\bf
  l}(\nabla\cdot\vec\theta_{\bf l}-q)^2$.  Alternatively one may shift
the $\vec\theta$ field by $ q{\bf l}$ to move the $q$--dependence into
the pinning term $\frac{\gamma(T)}{2\pi^2}\cos(2\pi(\theta_i+ql_i))$.
Since the grain coordinates $l_i$ take integer values, the model is
periodic in $q$--space with unit periodicity. Leaving
the analysis  of random background charges for future studies we will restrict
ourselves to uniform gate voltages $q({\bf l})=q=const.$ throughout.

For small $q$, the system is in a particle--hole symmetric ``neutral''
state: as for $q=0$, the ground state is given by $\vec\theta_{\bf
  l}=0$.  At some finite value $q=q^*$, a transition towards a charged
(with a non--integer average number of electrons per dot) and
spatially non--uniform ground state takes place.  To find $q^*$, let
us compute the soliton energy in the presence of $q$.  Since the
$q$--dependence of the Hamiltonian is a pure boundary effect, one
immediately finds the soliton energy $T^*(q)=T^*(0)-2qE_c$. At
$q^*=T^*(0)/(2E_c)$ the soliton energy $T^*(q)$ vanishes.  This marks
the transition into the charged state: for $q>q^*$, solitons are
created at no cost.

\begin{figure}[h]
  \centerline{\epsfxsize=2.25in\epsfbox{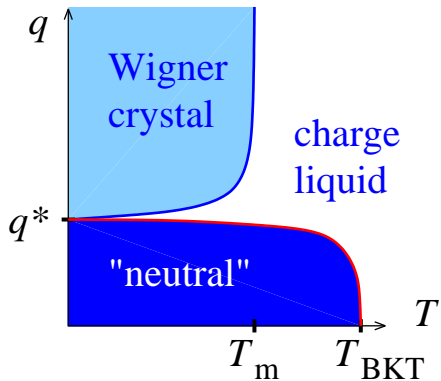}}
\caption{(Color online) The phase diagram of the 2{{d}} array with a uniform
background charge, $q$. In the ``neutral'' (insulating) phase, there are
(almost) no solitons. Outside the neutral regime,
charged solitons  are generated. Below the
melting temperature $T_{\rm m}$,  they form an insulating Wigner
crystal.} \label{fig8}
\end{figure}

It is instructive to relate the physics of the array to the phenomenon
of vortex formation in type II superconducting
films~\cite{deGennes}. Within this analogy~\cite{MKG}, the gate voltage
translates to an external magnetic field, $H$, and the gate voltage
$q^*$ corresponds to the critical magnetic field $H_{c1}$ where vortex
formation becomes  energetically favorable.

Above $H_{c1}$ a type II superconductor contains a finite density of
vortices which at low enough temperatures form an Abrikosov
lattice~\cite{abrikosov}. In a clean film, the vortex lattice is free
to move, but is easily pinned by the system boundaries, the underlying
lattice structure (as in Josephson junction arrays) or any sort of
disorder~\cite{Larkin,blatter}.  Upon increasing the temperature it
will eventually melt~\cite{BKT-kt,Tm}, and above the melting
temperature $T_{\rm m}$ most of the vortices are unbound.  The melting
temperature at finite $H>H_{c1}$ is smaller than the zero magnetic
field Berezinskii--Kosterlitz--Thouless but parametrically of the same
order~\cite{Tm}. Thus, at $T<T_{\rm m}$ the system is superconducting
while at $T>T_{\rm m}$ vortex motion leads to dissipation.

Translating back to our problem this means that at $q>q^*$ the
solitons form a Wigner crystal once their density is sufficiently
large such that the interaction is logarithmic. Only in the narrow
interval $q^*< q< q^* + \xi_{\rm s}^{-2}$ the system is in the
conducting charge liquid state. Upon increasing the gate voltage, the
Wigner crystal forms and, due to lattice pinning, the system is an
insulator at temperatures smaller than the melting temperature.  The
latter is of the order of $T_{\rm BKT}$. The phase diagram is shown in
Fig.~\ref{fig8}. Note that while for $q<q^*$
charge is carried by individual (thermally--activated) solitons, for
$q>q^*$ the mobile charges are lattice defects, whose core energy is
proportional to the logarithm of the lattice constant of the Wigner
crystal.

\subsection{DoS}

As in the 1{{d}} case,  interactions  lead to a suppression of the
tunneling DoS  in the vicinity of the Fermi energy.
Using Eq.~(\ref{eq-dos}), one finds that, in 2{{d}}, the suppression of
the DoS is determined by the phase correlator
\begin{eqnarray}
\sum_\bq\langle|\phi_\bq(\tau)-\phi_\bq(0)|^2\rangle\eq4T\sum_{\bq,m}
\frac{1-\cos\omega_m\tau}
  {|\omega_m|(\frac g\pi\bq^2+\frac{|\omega_m|}{E_c})}\nn\\
\eq\frac1{2\pi g}\ln^2(gE_c\tau).
\end{eqnarray}
The final expression for the DoS, thus, reads~\cite{tschersich,schwiete}
\begin{eqnarray}
\nu(\epsilon)\sim\nu_0\, \exp \left[-\frac1{4\pi g}
  \ln^2\frac{gE_c}\epsilon\right].
\end{eqnarray}
As in one dimension, the suppression of the DoS becomes significant at
a scale much larger than $T^*$, namely, at $\epsilon\sim
gE_ce^{-2\sqrt{\pi g}}$. While this is a purely perturbative result,
decreasing the energy further non--perturbative effects become
important. Specifically, the finiteness of the soliton energy leads to
a hard gap in the zero--temperature DoS,
\begin{equation}
\nu(\epsilon<T^*)\stackrel{T=0}{=}0.
\end{equation}


\section{Conclusions}
\label{sec-conclude}

We have studied transport properties of inelastic granular arrays. We
find that charge quantization and the localization of a unit charge
within a finite area play the crucial role in the low--temperature
behavior of these systems. This is the case even in systems where due
to a large bare conductance $g\gg 1$ charge behaves like a fluid at
high temperatures. At large $g$, the elementary charged excitations
are solitons spreading over (exponentially) many grains. In contrast,
for weakly coupled arrays ($g\ll 1$) charge is
quantized on a single grain. The excitation energy of the solitons, $T^\ast$,
determines the activation gap for the
low--temperature conductivity. In 2{{d}} arrays, logarithmic
interactions between solitons lead to a sharp BKT crossover to the
high--temperature regime at a temperature $T_{\rm BKT}$ which is
parametrically smaller than the activation gap: $T_{\rm
  BKT}=T^*_{(2{{d}})}/g\ll T^*_{(2{{d}})}$. The most straightforward way to
access this physics is to study the elementary excitations of the
classical, dissipative charge model. The latter may be viewed as
the low--frequency limit of the phenomenological charge model,
Eq.~(\ref{eq-matveev}), introduced in the context of a single
dot~\cite{flensberg,matveev}. The reduction to the low--frequency
(classical) sector proceeds through integrating out all the modes
with frequencies higher than a certain bandwidth (eventually to be
understood as the temperature.) This process renormalizes the
initial amplitude of the cosine--potential $E_c e^{-\cG_0/2}$,
bringing it down to $E_c e^{-\cG_0/2} (T/E_c)^{1-1/d}$. (The
renormalization is due to the massless rotational modes, which
rotate charge without compressing it. Such modes are absent in
1{{d}}, thus there is no renormalization for $d=1$.)  The
effective bandwidth reaches the amplitude of the
cosine--potential at a certain energy scale (freezing temperature)
$T_0$. The latter is determined by the condition:
\begin{equation}
\label{eq-Tzero}
\cG_0-{2\over d}\ln (E_c/T_0) =0 \,.
\end{equation}
It is reasonable to assume (based on the exact solution of
the single--dot model~\cite{matveev}) that the renormalization of
the potential  stops at this scale. The physical reason behind
this saturation is freezing--out of the rotational modes due to  mass
generation (by the cosine itself.) It is very important, however, to
realize that in 1{{d}} and 2{{d}} the soliton energy is {\em larger} than the
freezing temperature: $T^*\gg T_0$. As a result, the insulating
behavior is established independently of the validity of the freezing
assumption (the latter simply allows one to describe the insulator at
$T<T_0$.)

In this paper we have derived the above mentioned phenomenology
starting from the complimentary phase--model. The latter is
directly deducible from the microscopic fermionic Hamiltonian.
   The
mapping between the two approaches is achieved by summation over
instanton configurations (plus Gaussian fluctuations) of the
phase--model. Curiously, the non--Gaussian fluctuations of the phase
field around the instantons become strong at the same temperature
scale $T_0$ given by Eq.~(\ref{eq-Tzero}).  As was already mentioned
above, the insulating behavior in 1{{d}} and 2{{d}} sets in at the scale
$T^*$, which is parametrically larger than $T_0$ given by
Eq.~(\ref{eq-Tzero}). Thus, the instanton gas summation is fully
justified. In 3{{d}}, $T^*_{(3{{d}})}\approx T_0$, and therefore instantons
and non--Gaussian fluctuations become important at the very same
temperature, complicating the theoretical treatment. Our approach
allows one, thus, to follow the behavior of the (1{{d}} and 2{{d}}) system
from the metallic phase at high temperatures down to the insulating
phase at low temperatures.

However, this luxury comes at the price of several
key simplifications of the model. (i) The array is assumed to be
perfectly uniform, i.e., we choose grain capacitances, $C$,
inter--grain couplings, $g$, and (most restrictively) background
charges, $q$, to be identical for all the grains. (ii) We have assumed
a continuous spectrum in each grain. Doing this we have disregarded
all manifestations of quantum coherence, such as Anderson
localization, elastic hopping through several grains, etc. As a
result, our treatment misses the physics at energy scales associated
with the single--particle level spacing, $\delta$. Thus, it provides
only transient temperature dependencies --- and not the ultimate
low--temperature conductivity.

In principle, (random) fluctuations of the parameters ($C,\,g,\,q$)
may be incorporated in our treatment. They will directly translate
into static fluctuations of the corresponding parameters of the charge
model. If such fluctuations are smooth and long--range correlated,
solitons will continue to be the elementary charged excitations of the
model. The difference is that now the solitons move in the presence of
a random pinning potential. The problem is reduced to the dissipative
dynamics of a {\em classical} gas of interacting solitons and
anti-solitons subject to pinning. In 2{{d}} the very same model is
used to describe the motion of vortices in type II superconductors,
see, e.g., Refs.~[\onlinecite{blatter,doussal}]. In 1{{d}} the conductivity of
the array is determined by rare fluctuations causing anomalously
strong soliton pinning~\cite{foot2}.  Strong short--range disorder,
however, may invalidate the soliton picture.  It is not known whether
arrays with short-- and long--range disorder display the same
low--temperature behavior.

Incorporating quantum coherence is yet a more difficult task. To
proceed in this direction, one has to take into account the finite
level spacing $\delta$ in each grain. The most natural model assumes a
random chaotic spectrum in each grain with Wigner--Dyson spectral
statistics, mutually uncorrelated between different grains. Such a
model necessarily includes randomness, even if all the other
parameters are assumed to be strictly deterministic.  Technically, the
chaotic spectrum of a grain is described by the non--linear
$\sigma$--model matrix field $Q_{\bf l}$. Since our ultimate goal is
to describe the interacting system, either replica~\cite{Finkelstein}
or Keldysh~\cite{KamenevAndreev} variants of the latter should be
considered. In this case the $Q_{\bf l}$--field is a matrix in replica
(Keldysh) space as well as in energy space. One thus faces a theory of
two coupled fields: $Q_{\bf l}$ and $\phi_{\bf l}$. The perturbative
renormalization (RG) group treatment of such a theory is well
documented in the literature~\cite{Finkelstein,BelitzKirkpatrick}.  In
the absence of electron-electron interaction (no fluctuations of
$\phi_{\bf l}$), the electron states in a low--dimensional ($d=1,2$)
array are localized. Accounting for even small fluctuations of
$\phi_{\bf l}$ results in a finite phase coherence length. This length
gets shorter at higher temperatures, and, in the
case~\cite{vinokur2,blanter} of 
$d=2$, becomes of the order of the period of the granular array at $T\sim
g^2\delta$. At higher temperatures,  single-particle localization
is not important.  However, as has been shown in this paper, if
$\delta\ll E_c\exp[-g]$, the conductivity of an array still may be
strongly suppressed due to the formation of charge solitons. Those
appear due to the $\phi$--field configurations with non--zero winding
numbers, $W$, or instantons. The latter are absent in the perturbative
RG, making it inadequate for the quantitative description of the
insulating phase. It is known from the single--dot
model~\cite{Kamenev2000} that the $\phi$--field instantons induce
non--perturbative rotations of the $Q$--matrix fields. The full theory
accounting for the single-particle localiztion effect and for the
charge solitons, should therefore contain {\em combined} instantons of
$\phi$ and $Q$ degrees of freedom~\cite{andreev}.  The experience in a
non--perturbative treatment of replica (or Keldysh) non--linear
$\sigma$--models is relatively limited, but rapidly growing. We should
mention recent successful treatment of the Wigner--Dyson spectral
statistics~\cite{KamenevMezard,AltlandKamenev,Kanzieper}. Even more
encouraging is the recent realization that Anderson localization in
certain one--dimensional systems may be treated exactly by summing up
all stationary configurations of the $Q$--matric field with non--zero
winding numbers~\cite{LamSimZirn,AltKamTian}. These developments make
us optimistic that a quantitative treatment of the insulating phase of
disordered interacting electronic systems may be attained rather soon.


\begin{acknowledgments}
  We are grateful to K.~B.~Efetov, M. Fogler and A.~I. Larkin
  for valuable discussions.  Work at the University of Minnesota was
  supported by the A.P. Sloan foundation and the NSF grant
  DMR04-05212 (AK) and by NSF grants DMR02-37296, and EIA02-10736
  (LG).  JSM was partially supported by a Feodor Lynen fellowship
  of the Humboldt Foundation as well as by the U.S. Department of
  Energy, Office of Science, under Contract No.~W-31-109-ENG-38. Work
  at the University of Cologne was supported by Sonderforschungsbereich
  SFB/TR 12 of the Deutsche Forschungsgemeinschaft.
\end{acknowledgments}


\begin{appendix}

\section{Phase model}

\subsection{Derivation of the AES action}
\label{app-AES}

We start from a prototypical model
consisting of one dot (D) coupled to a lead (L). The system is
described by electronic degrees of freedom and its action is given by
\begin{eqnarray*}
S[\psi]\!\!\!&&=\,\,\,
  \sum_{i=D,L}S_i[\psi] + S_t[\psi] + S_c[\psi],
\end{eqnarray*}
where the non--interacting action $S_D$ ($S_L$) of the dot
(lead), a tunneling term $S_t$, and the charging action,  $S_c$, are
given by, respectively,
\begin{eqnarray*}
S_i[\psi]\eq \int_0^\beta d\tau\;
\bar \psi_i(\tau)\left(\partial_\tau - \mu + H_i\right)\psi_i(\tau),\\
S_t[\psi]\eq \int_0^\beta d\tau\left(
\bar \psi_D(\tau) \hat T \psi_L(\tau)+{\mbox{h.c.}}\right),\enspace{\rm and}\\
S_c[\psi]\eq E_c\int_0^\beta d\tau\;
(\hat \cN_D(\tau)-q)^2,
\end{eqnarray*}
and $\psi(\tau)$ is an imaginary--time fermionic (Grassmann) field.
Further,
\begin{itemize}
\item $H_D$ ($H_L$) is the Hamiltonian of the dot (lead). The notation
  $\bar \psi_i H_i \psi_i$ implies a sum over the internal Hilbert
  space, i.e., $\bar \psi_i H_i \psi_i = \int d\bx\; \bar\psi_i (\bx)
  H_i \psi_i(\bx)$.
\item $\hat T$ is the tunneling operator between the dot and the lead. Its
  real space representation is given by some matrix $T(\bx,\bx')$
  whose detailed structure we need not specify.
\item $E_c=e^2/(2C)$ is the charging energy, where $C$ is the
  self--capacitance of the dot, and $\hat \cN_D =\int d\bx\; \bar
  \psi_D(\bx)\psi_D(\bx)$ the number operator of the dot.
\item $q=V_{\rm g}C/e$ is the background charge on the dot set by an
  external gate voltage $V_{\rm g}$.
\end{itemize}
The decoupling of the interaction part of the action is effectuated by
the introduction of a Hubbard--Stratonovich field $V(\tau)$ with the
physical significance of a voltage:
$$
e^{-S_D[\psi]-S_c[\psi]} \longrightarrow \int {\cal D}V\;
e^{-S[V]-S_D[\psi,V]},
$$
where
\begin{eqnarray*}
S_c[V]\!\eq\!\frac1{4E_c}\int_0^\beta d\tau\; V^2(\tau)-iq\int_0^\beta
V(\tau),\\
S_D[\psi,V]\!\eq\! \int_0^\beta\!\! d\tau\; \bar
\psi_D(\tau)\left(\partial_\tau - \mu + H_D + i
  V(\tau)\right)\psi_D(\tau).
\end{eqnarray*}
The field $V$ can be removed from the dot action $S_D$ by the gauge
transformation
\begin{eqnarray}
\label{eq-gauge}
\psi_D(\tau)\to e^{i\phi(\tau)}\psi_D(\tau),\enspace{\mbox{where}}\enspace
\dot\phi(\tau)=V(\tau).
\end{eqnarray}
However, that transformation requires some caution. The fermionic
fields obey anti--periodic boundary conditions
$\psi_D(0)=-\psi_D(\beta)$. In order to preserve this property, the
field $\phi$ has to fulfill the condition $\phi(\beta)-\phi(0)=2\pi
W$, where $W\in\Bbb{Z}$. Therefore, the static contribution $V_0=\int
d\tau\; V(\tau)$ can only be removed up to $\delta V_0=V_0-2\pi T W$,
where $W=[V_0/(2\pi T)]$ is the closest integer to $V_0/(2\pi T)$.
However, in the limit of negligible level spacing $\delta\to 0$,
fluctuations around $\delta V_0=0$ are suppressed.

The gauge transformation (\ref{eq-gauge}) couples to the tunneling
terms
\begin{eqnarray}
S_t[\psi]\to S_t[\psi,\phi]=\! \int\limits_0^\beta\! d\tau\left(
\bar \psi_D(\tau) \hat T e^{i\phi(\tau)}\psi_L(\tau)+{\mbox{h.c.}}\right)\!.\nn
\end{eqnarray}
Since the action is quadratic, the fermionic fields can be easily
integrated out to yield a description in terms of the phase field
$\phi$ only.  Performing the Gaussian integration over $\psi$ we obtain
$$
S_t[\phi]=
-\tr\ln\left(\begin{matrix}G_D^{-1}&\hat Te^{i\phi}\\\hat T^\dagger
    e^{-i\phi}&G_L^{-1}\end{matrix} \right),
$$
where the trace extends over all index spaces (time, position,
$i=D,L$) and $G_i^{-1}=\partial_\tau-\mu+H_i$.

As a next step, we expand the '$\tr\ln$' in tunneling amplitudes,
\begin{eqnarray}
S_t[\phi]=\sum_k\frac1{2k}\tr\left[\left(G_D\hat Te^{i\phi}G_L\hat T^\dagger
    e^{-i\phi}\right)^k\right].
\end{eqnarray}
Finally, the trace over the Hilbert spaces of dot and lead,
respectively, leads to
\begin{equation}
\label{eq-AES0}
S_t[\phi] = -\frac18\sum_k\kappa_k \tr\left[\left(\Lambda
    e^{i\phi}\Lambda e^{-i\phi}\right)^k\right],
\end{equation}
where $\kappa_k=-4\frac{(-1)^k}k\sum_\alpha
(\pi^2\nu_D\nu_L|T_\alpha|^2)^k$, $\nu_D$ ($\nu_L$) is the density of
states  of the dot
(lead) and
\begin{eqnarray}
\Lambda(\tau-\tau')=- \frac{i}{\sin(\pi T(\tau-\tau'))}.
\end{eqnarray}
(In the main text, $\Lambda$ is given in Matsubara representation.)
Adding to this the phase representation of the charging action,
$S_c[\phi]=\int d\tau 
\,\dot\phi^2/(4E_c)-iq(\phi(\beta)-\phi(0))$, we obtain the phase
action of the system, $S=S_c + S_t$.

The straightforward  generalization of this result to
${{d}}$--dimensional array geometries is given by
Eqs.~(\ref{eq-S_c}), (\ref{eq-AES}).

\subsection{Fluctuation determinant}
\label{app-fluct_det}
In this appendix we provide details of the computation of the
determinants resulting from the integration over fluctuations around
instanton solutions. We will discuss the cases of isolated dots,
one-- and two--dimensional arrays in turn.

\subsubsection{Single quantum dot}
\label{sec:isolated-quantum-dot}

We begin by expanding the action (\ref{eq-S_0}) to
second order in deviations, $\delta\phi$, from $\phi_W(\{z\})$.  Since
the instantons are saddle point configurations, there are no linear
terms in this expansion:
\begin{eqnarray}
\delta S_{\rm inst}=g\langle\delta\phi|\hat
  F|\delta\phi\rangle,
\end{eqnarray}
where the operator $\hat F$ is given by
$$
\hat F(z)= \bullet \Lambda_W(z) \bullet \Lambda - {1\over 2}
\bullet\bullet \Lambda_W(z) \Lambda - {1\over 2}
\Lambda_W(z)\bullet\bullet \Lambda,
$$
and $\Lambda_W(z) \equiv e^{i \phi_W(z)}\Lambda e^{-i\phi_W(z)}$.
The dots indicate places for $\delta\phi(\tau)$ in the matrix
products. In a more explicit notation,
\begin{eqnarray}
2\hat F(\tau,\tau') \eq \Lambda_W^{\tau,\tau'}\Lambda^{\tau'\!-\!\tau} +
\Lambda^{\tau\!-\!\tau'}\Lambda_W^{\tau',\tau}\nn\\
&&-\delta(\tau\!-\!\tau')\!\int \!d\tau_1
\left[\Lambda_W^{\tau,\tau_1}\Lambda^{\tau_1\!-\!\tau} +
\Lambda^{\tau\!-\!\tau_1}\Lambda_W^{\tau_1,\tau}\right].\nn
\end{eqnarray}
For $z=0$, the operator $\hat F$ is diagonal in the Matsubara
frequency basis. Its spectrum is given by
$$
\lambda^{(W)}_m = \left\{ \begin{array}{ll} 0, &1\leq |m| \leq |W|,\\
    |m| - |W|,\enspace &|W|<|m|. \end{array}\right.
$$
For $z\neq0$, the eigenbasis has a more complicated form, but the
eigenvalues are independent of $z$. For $W=1$, the basis reads
\begin{eqnarray}
|\varphi_m(z)\rangle=\begin{cases}
{\displaystyle u^{m+1}\frac{1-uz^*}{u-z}} & m\leq-2,\\
{\displaystyle \sqrt{1-|z|^2}\frac1{u-z}} & m=-1,\\
{\displaystyle \sqrt{1-|z|^2}\frac u{1-uz^*}} & m=1,\\
{\displaystyle u^{m-1}\frac{u-z}{1-uz^*}} & m\geq2,\\
\end{cases}
\end{eqnarray}
where $u=\exp[2\pi i T\tau]$.
[$m=0$ corresponds to a constant shift which is of no interest.]
The quadratic action is thus given by
\begin{eqnarray}
\label{eq-Sf}
S_{\rm inst} = \cG_0|W| - 2\pi iWq+ g\sum_m \lambda^{(W)}_m
|\delta\phi_m|^2,
\end{eqnarray}
where $\delta\phi(\tau) = \sum_m\delta\phi_m |\varphi_m(z)\rangle$.

Now we can evaluate the partition function taking into account all
instanton configurations.  The partition function $Z$ is given by
Eq.~(\ref{eq-sum_W}),
\begin{eqnarray}
Z = Z_0\sum\limits_W \frac{Z_W}{Z_0} e^{2\pi i Wq},
\end{eqnarray}
where $Z_0$ is the partition function in the absence of instantons.
The contribution to the partition function $Z_W$ corresponding to a
certain winding number $W$ consists of all configurations with $s+W$
instantons and $s$ anti-instantons~\cite{Grabert96,fkls}, where
$s\geq\max\{0,-W\}$. Here we neglect the weak interaction between
(anti-)instantons (cf.~main text.) Thus,
\begin{eqnarray}
\frac{Z_W}{Z_0}=\!\!\!\sum_{s=\max\{0,-W\}}^\infty\!
\frac{(2s+W)!}{(s+W)!s!}\frac{{\cZ}_{2s+W}}{Z_0} \int (dz)\,{\cal
  J}^{(2s+W)}(z),\nn
\end{eqnarray}
where ${\cal J}^{(w)}(z)$ is the Jacobian of the transformation to the
collective coordinate basis $\{|m\rangle\} \to \{z,
|\varphi_{m>w}\rangle\}$,
$$
{\cal J}^{(w)}(z) = {1\over w!}\, \mbox{det}^{(w)} \left\|
  \frac{1}{1-z_a z^*_{a'}}\right\|.
$$
Furthermore, ${Z}_w$ is obtained by Gaussian integration over
the massive fluctuations $\delta\phi$, namely
\begin{eqnarray}
\frac{\cZ_w}{Z_0}\eq e^{-{\cG_0}w} \left(\frac{\prod_{m=1}^\infty
    g\lambda_m^{(0)}}
{\prod_{m=2}^\infty g\lambda_m^{(1)}}\right)^w
\!=\left(ge^{-{\cG_0}}\prod_{m>1}\frac m{m-1}\right)^w\!\!\!.\nn
\end{eqnarray}
Rewriting the product $\prod_m(\dots)=\exp[\sum_m\ln(\dots)]$, one
finds that it is dominated by large $|m|$. Therefore, it is sufficient
to keep only the first term in an expansion in $\frac1{|m|}$ --- an
upper cut--off is provided by the charging term in the action. Thus,
\begin{eqnarray}
\frac{\cZ_w}{Z_0}
&\simeq&  \left(g^2\frac{E_c}Te^{-\cG_0}\right)^w.
\end{eqnarray}
Finally,
\begin{eqnarray}
\int d^{2w}z\;\mbox{det}^{(w)} \left\| \frac{1}{1-z_a
z^*_{a'}}\right\|\sim\pi^w\ln^{w}\frac{E_c}T.
\end{eqnarray}
Putting all the components back together yields
\begin{eqnarray}
\frac {Z_W}{Z_0} =\sum_{s=0}^\infty
\frac{1}{(s+|W|)!\,s!}\left(\pi g^2\frac{E_c}Te^{-{\cG_0}}
  \ln\frac{E_c}T\right)^{2s+|W|}\enspace 
\end{eqnarray}
and, subsequently,
\begin{eqnarray}
\frac Z{Z_0} \eq\sum\limits_W e^{2\pi i
 Wq}I_{|W|}\left(2\pi g^2\frac{E_c}Te^{-{\cG_0}}\ln\frac{E_c}T\right)\nn\\
\eq\exp\left[2\pi g^2\frac{E_c}Te^{-{\cG_0}}\ln\frac{E_c}T\cos(2\pi q)\right],
\end{eqnarray}
where $I_\nu(z)$ is a Bessel function. This is the result quoted in
the main text.

\subsubsection{Fluctuation determinant of the 1{{d}} array}
\label{sec:fluct-determ-1d}

It is relatively straightforward to generalize the analysis of the
previous section to an array geometry. The main difference
to the single dot case is that the fluctuation matrix $F=\{F_{kl}\}$
has additional spatial structure, where the indices $k,l$ label the
dots in the array.  With the notation
\begin{eqnarray}
\lambda_m^{(k)}=
\begin{cases}
  |m| & |W_{k+1}-W_k|=0,\\
  |m|-1 \enspace & |W_{k+1}-W_k|=1,
\end{cases}
\end{eqnarray}
where $m$ is the Matsubara index, the fluctuation matrix $F$ for
the array takes the form
\begin{eqnarray}
F_{kk}=\lambda^{(k-1)}+\lambda^{(k)},\quad
F_{kk+1}=F_{k+1k}=-\lambda^{(k)},
\end{eqnarray}
and $F_{kl}=0$ otherwise. Here, we have put the instanton
parameter $z=0$. (This simplification is justified because the
dominant contribution to the fluctuation determinant comes from
high frequency fluctuations which are not significantly affected
by the value of $z$.)

It turns out to be convenient to rewrite $F=F_0-\delta F$, where $F_0$ is the
fluctuation matrix on the flat configuration without instantons,
$F_0^{kl}= |m|\left(2\delta_{k,l}-\delta_{k-1,l}-\delta_{
    k+1,l}\right)$. By contrast, $\delta F$ has non--zero entries
only at the edges of the plateau:
\begin{eqnarray}
\!\!\!\delta F_{kl}\eq \sum_{\{{
    s,s+1}\}\in\partial A}\Big(\delta_{k,l}(\delta_{
  k,s}+\delta_{k,s+1})\nn\\
&&\qquad\quad\qquad-\delta_{k,s}\delta_{
  l,s+1}-\delta_{k,s+1}\delta_{l,s}\Big),
\end{eqnarray}
where $\partial A$ is the boundary of the plateau, i.e.,
$\{{s,s+1}\}\in\partial A$ means that $|W_{s}-W_{s+1}|=1$.

Thus the fluctuation term ${\cal F}={\rm det}\,F/{\rm det}\, F_0$
can be represented as
\begin{eqnarray}
{\cal F}\eq\prod_m\exp\left[\tr\ln\left(|m|\tilde F_0-\delta
  F\right)-\tr\ln\left(|m|\tilde F_0\right)\right]\nn\\
\eq \exp\left[\sum_m \tr\ln\left(1 - \frac1{|m|}\tilde F_0^{-1}\delta
    F\right)\right],
\end{eqnarray}
where $F_0=|m|\tilde F_0$.

The Matsubara sum starts with $m=2$, i.e., there is no infrared
divergence.  The important contribution comes from large $|m|$. It is
therefore  sufficient to keep only the first term in
$\frac1{|m|}$, i.e.,
\begin{eqnarray}
\label{Fdet_freq_approx}
{\cal F}\simeq \exp\left[-\sum_m\frac1{|m|} \tr\left[\tilde F_0^{-1}\delta
    F\right]\right].
\end{eqnarray}
Inverting the matrix $\tilde F_0$ we find $(\tilde
F_0^{-1})_{kl}=\min\{k,l\}-\frac{kl}{M+1}$ which leads to
\begin{eqnarray}
\label{eq-A18}
\tr\left[\tilde F_0^{-1}\delta
  F\right]
\eq\frac{2M}{M+1}
=\begin{cases}1 & M=1,\\2 &M\to\infty.\end{cases}
\end{eqnarray}
We finally note that the summation over $m$ has to be cut off at
large frequencies $m\sim E_c/gT$ where the charging energy $\sim
E_c^{-1} \omega_m^2$ and the dissipation action $\sim g
|\omega_m|$ become comparable. This leads to the estimate $\sum_m
|m|^{-1} \sim \ln(g E_c/T)$. Substitution of this formula along
with Eq.~(\ref{eq-A18}) into Eq.~(\ref{Fdet_freq_approx}) leads to
the result \eqref{F_1D}.

\subsubsection{Fluctuation determinant of the 2{{d}} array}
\label{sec:fluct-determ-2d}

The two--dimensional fluctuation determinant may be obtained by
straightforward generalization of our discussion of the previous
section.
Concentrate for simplicity on the infinite system, $M\to\infty$, we
find that (the Fourier representation) of the
fluctuation matrix on the flat background (no instantons)
is given by $F_0 = |m| \tilde F_0$, where
\begin{eqnarray}
\label{eq-M02} \tilde F_0({\bf p})\eq
4\left(\sin^2\frac{p_x}2+\sin^2\frac{p_y}2\right).
\end{eqnarray}
To account for the presence of instantons, we again need to
evaluate the quantity $\tr\left[\tilde F_0^{-1}\delta F\right]$.
Since $\delta F$ has non--vanishing entries only at the boundary
of the island with $W=1$, one finds
\begin{eqnarray}
\tr\left[\tilde F_0^{-1}\delta
  F\right]\eq\!\!\!\sum_{\{{\bf
    s,s+e_i}\}\in\partial A}\left((\tilde F_0^{-1})_{\bf ss}+(
  \tilde F_0^{-1})_{\bf s+e_is+e_i}\right.\nn\\
&&\qquad\qquad\left.-(\tilde F_0^{-1})_{\bf ss+e_i}-(
  \tilde F_0^{-1})_{\bf s+e_is}\right),\nn
\end{eqnarray}
where $\{{\bf s,s+e_i}\}\in\partial A$ means that the link between
sites ${\bf s}$ and ${\bf s+e_i}$ crosses the boundary of the island,
i.e., $|W_{\bf s}-W_{\bf s+e_i}|=1$.  Using Eq.~(\ref{eq-M02}), one
obtains
\begin{eqnarray}
\tr\left[\tilde F_0^{-1}\delta
  F\right]\eq\sum_{\{{\bf
    s,s+e_i}\}\in\partial
  A}\sum_{\bf
  p}\frac{\sin^2\frac{p_i}2}{\sin^2\frac{p_x}2+\sin^2\frac{p_y}2}\nn\\
 \eq
\sum_{\{{\bf s,s+e_i}\}\in\partial
  A}\frac12=\frac L2,\nn
\end{eqnarray}
where $L$ is a number of links along the circumference of the
island.  Using the approximations \eqref{Fdet_freq_approx} and
$\sum_m |m|^{-1} \sim \ln(g E_C/T)$ (which apply regardless of
dimensionality) we then find that the fluctuation determinant is
given by Eq.~\eqref{Fdet_2D}.

\subsection{Conductance}
\label{app-conductance}

To compute the conductance, we couple the action to source terms $A_X$
($X=L,R$), 
\begin{eqnarray}
S_t\to \frac1{8}\sum_k\kappa_k\sum_{X=L,R}
\tr\!\left[\left(\Lambda
    e^{i(\phi\pm A_X)}\Lambda e^{-i(\phi\pm A_X)}\right)^k\right].\nn
\end{eqnarray}
Physically, these terms represent the vector potential of an electric
field coupled to the system.  The Kubo conductance, $G$, may be
computed by taking two--fold derivatives of the partition function
with respect to $A_X$ at $A_X=0$:
\begin{eqnarray}
\label{eq-Gapp}
 G=-\frac\pi2\lim_{\omega\to
  0}\frac T\omega\Im\left[Q(i\omega_m)\right]_{i\omega_m\to\omega^+},
\end{eqnarray}
where
\begin{eqnarray}
Q(i\omega_m)=\frac1Z\left.\frac{\delta^2 Z}{\delta A_R(\omega_m)\delta
    A_L(-\omega_m)}\right|_{A_R=A_L=0}. 
\end{eqnarray}
Representing the partition function as a sum over winding number
sectors, one finds
\begin{eqnarray}
Q(i\omega_m)=\frac1Z\sum_W e^{2\pi iWq}  \left\langle\frac{\delta
S_W}{\delta A_R(\omega_m)}\frac{\delta S_W}{\delta
A_L(-\omega_m)}\right\rangle .\nn
\end{eqnarray}
Or, taking into account instantons and anti-instantons,
\begin{eqnarray}
\label{eq-Q}
\!\!\!Q(i\omega_m)=\frac1Z\!\sum_W e^{2\pi iWq}\!
\sum_{s=0}^\infty\! \frac{(2s+|W|)!}{(s+|W|)!s!}q^{(2s+|W|)}_m\!,
\end{eqnarray}
where
\begin{eqnarray}
q^{(w)}_m=4g^2\cZ_w\int
d^{2w}z\;{\cal J}^{(w)}(z)\left\langle|\langle\delta\phi|\hat
  F_w|m\rangle|^2\right\rangle_{\delta\phi}.\quad 
\end{eqnarray}
Furthermore,
\begin{eqnarray}
&&\left\langle|\langle\delta\phi|\hat
  F_w|m\rangle|^2\right\rangle_{\delta\phi}\nn\\ 
\eq\sum_{k,k'}\lambda_l\lambda_{k'}\langle m|\varphi_k(z)\rangle
\langle\varphi_{k'}(z)|m\rangle
\left\langle\delta\phi_k\delta\phi_{-k'}\right\rangle_{\delta\phi}\nn 
\end{eqnarray}
and, according to the action Eq.~(\ref{eq-Sf}),
$\langle\delta\phi_k\delta\phi_{-k'}\rangle=\cZ_w\delta_{kk'}/(2g\lambda_k)$.
Thus,
\begin{eqnarray}
q^{(w)}_m\eq2g\cZ_w\int
d^{2w}z\;{\cal J}^{(w)}(z)\langle m|\hat F_w(z)|m\rangle.\nn
\end{eqnarray}
For small $m$, one finds
\begin{eqnarray}
\!\!\!\langle m|\hat F_{w}(z)|m\rangle\simeq
m\left(1+\sum_{j=1}^{w}\ln |z_j|^2\right)\!+{\cal O}(m^2).
\end{eqnarray}
Finally, using that
\begin{eqnarray}
{\cal J}^{(w)}(z)\simeq\frac1{w!}\prod_{j=1}^{w}\frac1{1-|z_j|^2},
\end{eqnarray}
we obtain
\begin{eqnarray}
q^{(w)}_m
&\simeq&2g\frac m{w!}{\cZ}_w\int
d^{2w}z\;(1+\sum_{j=1}^{w}\ln
|z_j|^2)\prod_{j=1}^{w}\frac1{1-|z_j|^2}\nn\\
\eq2g\frac m{w!}{\cZ}_w\left(\pi \ln\frac{E_c}T\right)^{w}
\left\{1+Aw\frac1{\pi \ln\frac{E_c}T}\right\},
\end{eqnarray}
where $A=-\pi^3/6$.

Substituting these results into Eq.~(\ref{eq-Q}) we obtain
\begin{widetext}
\begin{eqnarray}
Q(i\omega_m)\eq 2mg\frac{Z_0}Z\sum_W e^{2\pi
  iWq}\sum_{s=0}^\infty\!\frac{1}{(s+|W|)!s!}\left(g^2\frac{E_c}Te^{-{\cG_0}}
    \ln\frac{E_c}T\right)^{2s+|W|} \left\{1+A(2s+|W|)
    \frac1{\ln\frac{E_c}T}\right\}\nn\\  
\eq 2mg(1+2Ag^2\frac{E_c}Te^{-{\cG_0}}\cos(2\pi q)).\nn
\end{eqnarray}
\end{widetext}
Recalling that $\omega_m=2\pi m T$ and substituting in
Eq.~(\ref{eq-Gapp}), one finds Eq.~(\ref{eq-0d_cond}) for the
conductance.

\subsection{Coulomb gas mapping}
\label{app-SG}

We briefly review the mapping (1{{d}} Coulomb gas)$\leftrightarrow$ (1{{d}}
discrete sine--Gordon model)~\cite{map}. The action of the latter is
given by
\begin{eqnarray}
\label{eq-CG}
S[\theta]\eq\frac{E_c}T\int
dx\left\{(\partial\theta)^2 - 2ge^{-\cG_0/2} \cos(2\pi\theta)\right\}\!\!,\quad
\end{eqnarray}
where we have used specific expressions for the charge and the
fugacity
matching our model parameters (cf.~Eqs.~(\ref{eq-SG1}) and (\ref{eq-CGres}).)

Starting from the Coulomb gas representation, Eq.~(\ref{eq-CG0}),
we introduce the density variable
$\rho(x)=\sum_i\sigma_i\delta(x\!-\!x_i)$. Using the equality $1=\int
D\rho\; \delta(\rho(x)-\sum_i\sigma_i\delta(x\!-\!x_i))=\int
D\tilde\theta \, D\rho\;\exp[i\tilde\theta(x)
(\rho(x)-\sum_i\sigma_i\delta(x\!-\!x_i))]$, we may rewrite the
interaction term of the Coulomb gas in terms of the new variables and
subsequently integrate out $\rho$. This obtains a representation in
terms of the Lagrange multiplier field $\tilde\theta$,
\begin{eqnarray}
&&\prod_{i=1}^{2k}\int\limits_0^M dx_i\;
e^{\frac{\pi^2T}{E_c} \sum_{j,j'=1}^{2k}
  |x_j-x_{j'}|\sigma_j\sigma_{j'}} \nn\\
\eq \int D\tilde\theta \;e^{-\frac{E_c}{4\pi^2T}\int dx\;
  (\partial\tilde\theta)^2}\left|\int dx\;
  e^{i\tilde\theta(x)}\right|^{k}\left|\int dx\;
  e^{-i\tilde\theta(x)}\right|^{k}\!\!.\nn
\end{eqnarray}
Instead of imposing strict charge neutrality, one may assume that
positive and negative charges fluctuate independently,
\begin{eqnarray}
\!\!\!&&\!\!\!\sum_{k=1}^\infty\left(e^{-\cG(T)/2}\right)^{2k}
\frac1{(k!)^2} \left|\int dx\;
  e^{i\tilde\theta(x)}\right|^{k}\left|\int dx\;
  e^{-i\tilde\theta(x)}\right|^{k}\nn\\
  &\to&\prod_{\sigma=\pm1} \sum_{k=1}^\infty\frac1{k!}
  \left(e^{-\frac{\cG(T)}2}\int
    dx\;e^{i\sigma\tilde\theta(x)}\right)^k. 
  \end{eqnarray}
Performing the $k$--summations and relabeling
$\tilde\theta\to2\pi\theta$, we obtain $Z=\int D\theta
\,\exp(-S[\theta])$, where the action is given by \eqref{eq-CG}.


\section{Charge model}

\subsection{Multi--channel contacts}
\label{app-gamma}

In this appendix we discuss the generalization of the classical model
derived in the main text (Sec.~\ref{subsec-charge})  to $N\geq2$
channels.  For every channel $\alpha=1\dots N$ of a multi--channel
contact, a field $\theta_\alpha(\tau)$ is introduced.  In order to
clarify the following evaluation scheme, the reflection coefficients
$r$ are given indices specifying the direction, contact, and
channel. For simplicity, we consider the case of zero gate voltage
$q=0$ throughout. The quadratic action of a ${{d}}$--dimensional array
of $M^{{d}}$ grains 
with $N$ channels in each of the $dM^{{d}}$ contacts is then given by
\begin{eqnarray}
S_2 = \frac1T\sum_{{\bf l},m}\left\{\pi|\omega_m|\sum_\alpha
    \vec\theta_{{\bf l},\alpha}^{\;2}+E_c\Big(\sum_\alpha
    \nabla\cdot\vec\theta_{{\bf l},\alpha}\Big)^2\right\}\!,
\end{eqnarray}
while the backscattering is described by
\begin{eqnarray}
S_r = -\frac{E_c}\pi\sum_{{\bf l},\alpha} \sum_{i}r_{i{\bf
    l}\alpha}\! \int\limits_0^\beta\! d\tau\,
  \cos(2\pi\theta_{i,{\bf l},\alpha}).
\end{eqnarray}
Here, it is assumed that the high--energy modes $E_c<|\omega_m|<D$
have already been integrated out. [At energies larger than $E_c$, all
modes are decoupled and, thus, can be integrated out for each channel
separately.]

Only the ${{d}}M^{{d}}$ symmetric modes $\theta_{\bf
  l}=\sum_\alpha\theta_{{\bf l},\alpha}$ couple to external
parameters, such as gate voltages. We thus need to find an effective
action for $\theta_{\bf l}$ by integrating out the
$dM^d(N-1)$ asymmetric modes. To this end let us change variables from
$\theta_{{\bf l},\alpha}$ ($\alpha=1\dots N$) to $\theta_{\bf l}$ and
$\tilde \theta_{{\bf l},\alpha}=\theta_{{\bf l},\alpha}-(\theta_{\bf
  l}-\sum_{\alpha'>\alpha}\theta_{{\bf l},\alpha'})/(\alpha+1)$
($\alpha,\alpha'=1\dots N-1$). While the charging term renders the
symmetric fields $\theta_{\bf l}$ massive, all the asymmetric fields
$\tilde \theta_{{\bf l},\alpha}$ are massless. As a result, in the
perturbation theory in powers of $r_{i{\bf l}\alpha}$, terms
containing the massless fields $\tilde \theta_{{\bf l},\alpha}$ in the
exponents (cosines) vanish.  Rewriting the backscattering action in
terms of the new fields, one can see that the lowest order
non--vanishing terms are of the order $\prod_{\alpha=1}^Nr_{i{\bf
    l}\alpha}$, where the product runs over {\em all} channels of a
given contact:
\begin{widetext}
\begin{eqnarray}
Z_N\sim E_c^N\prod_{\alpha=1}^Nr_{i{\bf l}\alpha}\int
d\tau_\alpha\,\cos\left(\frac{2\pi}N\sum_\alpha \theta_{i,{\bf
      l}}(\tau_\alpha)\right) \prod_{\alpha=1}^N \left\langle
  \exp\left\{2\pi
  i\left(\tilde\theta_{i,{\bf l}\alpha}(\tau_\alpha) -
    \frac1\alpha\left(\tilde\theta_{i,{\bf l},\alpha}(\tau_N) +
      \sum_{\alpha'<\alpha}\tilde\theta_{i,{\bf
          l},\alpha}(\tau_{\alpha'}) \right)\right)\right\}\right
\rangle_{\tilde\theta_{{\bf l},\alpha}}\!\!\!.\nonumber
\end{eqnarray}
\end{widetext}
Taking the averages $\langle\dots\rangle_{\tilde\theta_{{\bf
      l},\alpha}}$ with the actions
\begin{eqnarray}
S_\alpha[\theta_{i,{\bf
    l},\alpha}] = \frac1T \sum_m \frac{\alpha+1}\alpha\pi
|\omega_m|\tilde\theta_{i,{\bf
    l},\alpha}^2\, ,
\end{eqnarray}
we obtain $E_c^{1-N}\prod_{\alpha=1}^N \prod_{\alpha'>\alpha}
(\tau_\alpha-\tau_{\alpha'})^{-2/N}$ for the product of correlators
$\prod_\alpha\langle\dots\rangle_{\tilde\theta_{{\bf l},\alpha}}$.
The re--exponentiated action of $\vec\theta_{\bf l}$ thus assumes the
time non--local form
\begin{widetext}
\begin{eqnarray}
S[\vec\theta]\!=\!\sum_{\bf l}\left\{\!\frac1T\sum_m\left( \frac\pi N
    |\omega_m| \vec\theta_{{\bf l}}^{\;2}+E_c\Big(
    \nabla\!\cdot\!\vec\theta_{\bf l}\Big)^2\right) -
  \frac{E_c}\pi\! \sum_{i}
  \prod_{\alpha=1}^N r_{i{\bf l}\alpha}\!\! \int\! d\tau_\alpha
  \prod_{\alpha'>\alpha}
  \frac1{(\tau_\alpha\!-\!\tau_{\alpha'})^{2/N}}
  \cos\Big(\frac{2\pi}N \sum_\alpha\theta_{i,{\bf
      l}}(\tau_\alpha)\Big) \!\right\}.
\end{eqnarray}
\end{widetext}
As a next step, in analogy to the single--channel case (see
Secs.~\ref{sec-1d_charge} and \ref{sec-2d_charge}), one may integrate
out all the remaining modes except the static one, $\theta_{m=0}$. The
prefactor of the cosine--term $E_c\prod_{\alpha=1}^N
r_\alpha\equiv \frac1{2\pi} E_c\gamma_0$ is renormalized according to
\begin{eqnarray}
\gamma_0&\to&\gamma(T)=\gamma_0\,\exp\!\Big\{-\frac{2\pi^2}{N^2}
\sum_{\alpha,\alpha'} \langle
\theta(\tau_\alpha)\theta(\tau_{\alpha'})
\rangle_{\theta_{m\neq0}}\Big\}\nonumber\\
&=& \gamma_0\,\exp\!\Big\{-\!\sum_{m\neq0} f(\omega_m)\big(1 \!+\!
\frac2N\sum_{\alpha,\alpha'>\alpha}
\cos\omega_m\tau_{\alpha\alpha'}\big)\Big\},\nonumber
\end{eqnarray}
where
\begin{eqnarray}
f(\omega_m)=2\frac
T{(2M)^d}\sum_\bq\left\{ \frac{\frac 1d}{NE_\bq+\pi|\omega_m|} +
  \frac{1-\frac1d}{\pi|\omega_m|} \right\}
\end{eqnarray}
and $E_\bq=4E_c\sum_i\sin^2(\pi q_i/(2M))$.

Since typical time differences
$\tau_{\alpha\alpha'}=\tau_\alpha-\tau_{\alpha'}$ are of the order
$1/T$ (the time integrals are dominated by the upper limit of
integration), the cosine--term inside the exponent may be
disregarded. Performing the Matsubara summation, one thus finds
$\gamma(T)=\gamma_0(T/E_c)^{1-1/d}$.

Evaluating the multiple time integrations in the prefactor of the
cosine, one finds that the integral over the center--of--mass time
$\tau=\sum_\alpha\tau_\alpha/N$ contributes a factor $1/T$, while the
integration over $N-1$ independent time differences
$\tau_{\alpha}-\tau$ yields a constant $c_N$ multiplied by the
logarithmic factor~\cite{ABG} $\ln E_c/T$ (following simply from power
counting.) The same logarithmic factor appears in the framework of the
phase model as a result of zero--mode integration. We shall not keep
this logarithm explicitly because all our evaluations of $\gamma$ are
done up to a numerical factor only.  As a result of these
approximations, we reproduce the classical model with $\gamma(T)\sim
(T/E_c)^{1-1/d}\, \prod_{\alpha=1}^N r_\alpha=
(T/E_c)^{1-1/d}\,e^{-\cG_0/2}$.

\subsection{Classical dynamics of the charge model}
\label{app-Keldysh}

Our strategy in dealing with the charge model was to eliminate all
 high--frequency degrees of freedom until only the zero
Matsubara frequency remains. The resulting theory describes the
classical statistical mechanics of interacting charges. This is
perfectly sufficient to describe the thermodynamics of the system.
To calculate the conductivity, however, one needs to retain the
low--frequency ($\omega \ll T$) dynamics. The Matsubara formalism
is not convenient for this purpose. Thus, we adopt the following
strategy: we first switch to the Keldysh formulation and then
integrate out the high--frequency components, reducing the theory
to the classical, $\omega\ll T$, sector only.

For the sake of simplicity, we deal here with the 1{{d}}
single--channel model, defined by Eq.~(\ref{eq-matveev}). In the
Keldysh formulation, the imaginary--time field $\theta(\tau)$ is
substituted by the pair of  {\em real--time} fields
$\theta^{cl}(t)$ and $\theta^{q}(t)$, originating from the
symmetric and antisymmetric combinations of the two branches of
the Keldysh contour, respectively. The operator in the quadratic
part of the action: $\pi|\omega_m|-E_c \nabla^2$ in
Eq.~(\ref{eq-matveev}) is transformed into a $2\times 2$ matrix in
$cl-q$ space. According to the general structure of the Keldysh
technique~\cite{Kamenev01}, the $cl-cl$ element of this matrix
must vanish (reflecting the fact that a field, purely symmetric on
the two branches of the contour, has zero action.) The $cl-q$ and
$q-cl$ elements are the retarded and advanced analytical
continuations of the Matsubara operator: $\pm\pi i \omega-E_c
\nabla^2\to \pm\pi\partial_t - E_c\nabla^2$. Finally, the $q-q$
element is given by the fluctuation--dissipation theorem (in
equilibrium): $2i\omega\coth(\omega/2T)$. In the classical limit,
$\omega\ll T$, it reduces to $4iT$.

Integration over the high--frequency components of the fields
leads to a renormalization of the backscattering amplitude, in
exactly the same way as in Matsubara theory. The remaining
renormalized backscattering part of the action, $\cos(2\pi\theta)$,
transforms to
$\cos(2\pi(\theta^{cl}+\theta^{q}))-\cos(2\pi(\theta^{cl}-\theta^{q}))\approx
-4\pi\theta^q\sin(2\pi\theta^{cl})+O((\theta^q)^3)$. Anticipating
that the quantum component fluctuates only weakly in the classical
limit, we omit  higher order terms in this expansion. As  a
result, the low--frequency part of the Keldysh action takes the form
\begin{eqnarray}\label{eq-keldcharge}
  S&=&\!\!\int\!dt\, \sum\limits_j  \Big[
  2\theta^q_j\left(\pi\partial_t\theta^{cl}_j-E_c\nabla^2\theta^{cl}_j
    +{E_c\gamma \over 2\pi}\sin(2\pi \theta^{cl}_j)\right)  \nonumber
  \\ 
  &&\qquad\qquad+ 4\pi iT\left(\theta^{q}_j\right)^2\Big] \, .
\end{eqnarray}
One can transform the last term in this action, employing a
Hubbard--Stratonovich transformation to the partition function $Z=\int
D\theta^{cl}\,D\theta^q\;\exp[iS]$, as 
\begin{equation}\label{eq-noiseHS}
  e^{-4\pi T \int dt\, \left(\theta^{q}_j\right)^2} = \int\!{\cal D}\xi
  \; e^{-\int\! dt\left\{{1\over 4\pi T} \xi_j^2 - 2i
      \xi_j(t)\theta^{q}_j(t)\right\}}\, . 
\end{equation}
Substituting this expression into Eq.~(\ref{eq-keldcharge}), one
notices that the resulting action depends on $\theta^q_j(t)$
only linearly. As a result, the integration over this field leads
to a functional $\delta$--function, imposing the following identity
at every instance of time:
\begin{equation}\label{eq-langkeld}
\pi\partial_t\theta^{cl}_j - E_c\nabla^2 \theta^{cl}_j
+{E_c\gamma \over 2\pi}\sin(2\pi \theta^{cl}_j) - \xi_j(t)=0\, .
\end{equation}
This is the Langevin equation, where $\xi_j(t)$ is the
noise with the correlator that may be read out from
Eq.~(\ref{eq-noiseHS}), namely $\langle \xi_j(t)\xi_{j'}(t')\rangle =
2\pi T \delta_{jj'}\delta(t-t')$. 

Finally, in the multi--channel case one must renormalize the
backscattering amplitude $\gamma$, as discussed in
appendix~\ref{app-gamma} above. In addition, the coefficient in front
of the time derivative must read $\pi/g$. This may be shown
phenomenologically by requiring that without backscattering
Eq.~(\ref{eq-langkeld}) describes the dynamics of the classical
$RC$--circuit. From the bosonization perspective, this coefficient is
given by $\pi/N$, where $N$ is number of channels, simply because in
the weak backscattering limit $g\approx N$. As a result, one recovers
Eqs.~(\ref{eq-langevin}) and (\ref{eq-noise}), used in the main text.

\end{appendix}



\begin{thebibliography}{}
\bibitem{foot1} Omission of the Gaussian fluctuations in  early
 treatments of the instanton gas~\cite{fs} lead to the erroneous
 conclusion that there is a quantum metal--insulator transition at
 $g_c\approx 1$ in 2d.
\bibitem{tunneling} If all $T_\alpha$ are small, $S_t$ can be
  simplified by retaining only $\kappa_1\simeq g$, where $g$ is the
  dimensionless (tunneling) conductance of the contacts. In this
  regime,
\begin{equation}
\label{eq-AES_g}
S_t[\phi]=\frac{g T^2}4\sum_{\langle\bl,\bl'\rangle}\int d\tau \, d\tau'\,
\frac{ \sin^2
  \left(\frac12(\phi_{\bl\bl'}(\tau)-\phi_{\bl\bl'}(\tau'))\right)}
{\sin^2(\pi T(\tau-\tau'))}. \nn
\end{equation}
\end{thebibliography}
\end{document}